\documentclass[a4paper,10pt]{article}
\usepackage[utf8x]{inputenc}
\usepackage{epsfig}
\usepackage{graphicx}
\usepackage{amsmath}
\usepackage{amssymb}
\usepackage{cite}
\usepackage{fancyhdr}
\newcommand{\lsim}{\raisebox{-0.13cm}{~\shortstack{$<$ \\[-0.07cm] $\sim$}}~} 
\newcommand{\gsim}{\raisebox{-0.13cm}{~\shortstack{$>$ \\[-0.07cm] $\sim$}}~} 

\title{$W$ mass and Leptonic $Z$-decays in the NMSSM}
\author{Florian Domingo\footnote{email: domingo@particle.uni-karlsruhe.de} \ and Teresa Lenz\vspace{2mm}\\ 
\em Institut f\"ur Theoretische Teilchenphysik\\
\em Karlsruher Institut f\"ur Technologie (Universit\"at Karlsruhe)\\
\em D-76128 Karlsruhe, Germany}
\date{}
\begin{document}
\maketitle
\thispagestyle{fancy}
\fancyhead[R]{TTP11-01\\SFB/CPP-11-03}

\begin{abstract}
We study a subset of electroweak-precision observables consisting of $M_W$, $\sin^2\theta_{\mbox{\tiny eff}}^{
\tau}$, $BR(Z\to\tau^+\tau^-)$ and $\Gamma(Z\to\tau^+\tau^-)/\Gamma(Z\to e^+e^-)-1$ (characterizing leptonic 
$Z$-decays) in the context of the NMSSM. After a brief review of common MSSM-NMSSM effects ({\em e.g.} for 
$\Gamma(Z\to\tau^+\tau^-)/\Gamma(Z\to e^+e^-)-1$, which has been little discussed, even in the MSSM), specific 
NMSSM scenarios are studied, with the result that the NMSSM, considering existing constraints on its spectrum, 
is essentially consistent with available measurements, given the current accuracy.
\end{abstract}

\section{Introduction}
ElectroWeak Precision Observables (EWPOs) \cite{EWPO} are sensitive to quantum corrections and thus to possible 
new physics effects. They are very accurately measured: current experimental data from LEP, SLD \cite{LEP_SLD} 
or TeVatron \cite{TeVatron} reach percent-level accuracy. Future improvements are expected from LHC \cite{LHC} 
and, in a more distant future, from a possible ILC with GigaZ option \cite{GigaZ}. Such a high level of 
precision makes these EWPOs useful to probe the Standard Model (SM) and its extensions, and constrain, provided 
the same accuracy can be matched on the theoretical side, the parameter spaces of such models. They are, in any 
case, unavoidable tests of the electroweak symmetry-breaking sector.

In the SM, the EWPOs are currently used to constrain the only unknown free parameter, {\em i.e.} the Higgs mass, 
leading to the result $M_H^{SM}=76^{+33}_{-44}$~GeV ($M_H^{SM}\leq144$~GeV at $95\%$ confidence level) 
\cite{MHSMfit} (see also \cite{LEP_SLD,TeVatron}). One observes the slight tension between this range and the 
current experimental lower bound on $M_H^{SM}$, resulting from direct searches at LEP ($M_H^{SM}\gsim114$~GeV) 
\cite{LEPHSM}. Beyond the SM, its simplest (softly-broken) supersymmetric extension, the MSSM, was already 
thoroughly investigated under the light of EWPOs (see 
\cite{Heinemeyer:2004gx,Heinemeyer:2006px,Heinemeyer:2007bw} for comprehensive reviews on the subject): several 
new contributions are expected, both from the extended Higgs sector and from the supersymmetric sector. Yet, a 
preference for rather light SUSY spectra (in the mSUGRA version of the MSSM) has been noted \cite{MSSMscale}, 
while, on the other hand, sensibly heavy SUSY particles are required to generate sufficiently large corrections 
to the lightest Higgs mass (in order to circumvent the experimental lower bound) \cite{LEPHMSSM}.

In the following, we will focus on the popular singlet ($\mathbb{Z}_3$-invariant, minimally CP and 
flavour-violating) extension of the MSSM, known as the Next-to-Minimal Supersymmetric Standard Model (NMSSM; 
see \cite{Ellwanger:2009dp} for a review and standard notations for NMSSM parameters). The leading motivation 
behind this singlet extension lies in solving the $\mu$-problem of the MSSM \cite{Kim:1983dt}. Moreover, 
several remarkable Higgs scenarii are known to be viable in the NMSSM, resulting in {\em e.g.} possibly light 
CP-even states $h_1$, either dominantly singlet (hence having reduced coupling to the SM sector) 
\cite{Ellwanger:1999ji} or decaying unconventionally into light CP-odd Higgs $A_1$ below the $B-\bar{B}$ 
threshold (see {\em e.g.} \cite{Dermisek:2005ar}). This last possibility has been further constrained by a 
new-analysis of {\sc Aleph} data \cite{ALEPH:2010aw}. Nevertheless, reduced branching ratios of the $A_1$ into 
$\tau^+\tau^-$, according to \cite{Dermisek:2010mg}, or increased singlet components of the $h_1$, may still be 
consistent with experimental limits. Significant interest was dedicated to $m_{h_1}\sim98$~GeV 
\cite{Dermisek:2005gg}, where, in virtue of its reduced couplings or its unconventional decays, a NMSSM CP-even 
Higgs could explain the small excess ($2.3\,\sigma$) observed at LEP in the channel $e^+e^-\to Z+b\bar{b}$ 
\cite{LEPHMSSM}. Note also that, independently of the CP-even Higgs (with $h_1$ possibly as heavy as 
$\sim140$~GeV in the NMSSM), light CP-odd Higgs represent another original NMSSM possibility with respect to 
the MSSM and may be natural in a R or Peccei-Quinn symmetry limit of the parameter space. Other NMSSM-specific 
effects are known to develop for large values of the parameter $\lambda$, such as increased CP-even Higgs 
masses (making low values of $\tan\beta\sim1-2$ acceptable) or reduced charged Higgs masses.

EWPOs were already partially discussed in the NMSSM: \cite{Cao:2008rc} dedicates particular interest to the 
$Z\to b\bar{b}$ channel, where the experimental result shows a discrepancy of about $\sim3\,\sigma$ with the SM 
prediction, with the conclusion that this feature could not be improved in the NMSSM. \cite{Cao:2010na} studied 
$Z$-decays involving very-light CP-odd Higgs bosons, {\em e.g.} in the NMSSM, in view of future searches at 
GigaZ. Yet, the details of the NMSSM phenomenology in view of the EWPOs, and especially in its specific 
scenarii, has been little discussed so far. This paper focuses therefore, in the context of the NMSSM, on the 
relation of the $W$ mass $M_W$ to the other parameters of the gauge sector: the $Z$-mass $M_Z$, the Fermi 
constant $G_{\mu}$ (obtained from muon decays) and the fine-structure constant $\alpha$. Leptonic $Z$-decays 
in the NMSSM are also discussed. We find that most of the NMSSM, in its allowed parameter space (in view of 
stability conditions, phenomenological constraints from LEP, $B$- or $\Upsilon$-physics,\ldots), is consistent 
with existing bounds on $M_W$ and leptonic $Z$-decays. The leading new-physics effects are associated with the 
supersymmetric sector (provided the SUSY spectrum is rather light), hence MSSM-like. NMSSM-specific effects 
associated with the Higgs sector are essentially determined by the mass of the light doublet state and 
Standard-Model-like. In the following section, we detail our calculation of $M_W$ and leptonic $Z$-decays in 
the context of the NMSSM. In the third section, we will discuss our results for the NMSSM, with particular 
emphasis on NMSSM-specific scenarii. In the remaining part of the introduction, we shall summarize the current 
experimental and SM status of those observables that we plan to discuss.

The current experimental value of $M_W^{\mbox{\tiny exp.}}=80.399\pm0.023$~GeV originates from a combined fit 
of LEP and TeVatron measurements \cite{LEP_SLD,TeVatron}. Future improvements should reduce the error down to 
$15$~MeV \cite{LHC} and $7$~MeV \cite{GigaZ} at LHC and GigaZ respectively. In the SM, one-loop 
\cite{MWSM_1loop} and two-loop \cite{MWSM_2loop,Awramik:2003rn} corrections as well as leading higher order 
effects (to the parameter $\Delta\rho$) \cite{Drho_highord} have been computed, leading to an estimated 
theoretical error of about $4$~MeV \cite{Awramik:2003rn}. Several computer codes (TOPAZO \cite{TOPAZO}, ZFitter 
\cite{ZFitter1,ZFitter2}, GFitter \cite{GFitter}) evaluate this quantity, relying on fitting expressions such 
as eq. 5.2-5.4 in \cite{ZFitter2}. The uncertainties on the top mass $m_t=172\pm1.6$~GeV, $M_Z=91.1876\pm0.0021
$~GeV, $\alpha_S(M_Z)=0.1184\pm0.0004$ and the fermionic shift of the fine structure constant 
$\Delta\alpha=0.05907\pm0.0004$ \cite{PDG} yield an additional parametric error, which we evaluate by varying 
these quantities within eq. 5.2-5.4 of \cite{ZFitter2}. Assuming (arbitrarily) a SM Higgs mass of 
$M_H^{SM}=150$~GeV (as we will all along this paper, except in plots, when otherwise mentioned), we eventually 
come to a SM estimate: $M_W^{SM}(M_H^{SM}=150~\mbox{\small GeV})=80.343\pm(0.012)^{\mbox{\tiny param.}}\pm(
0.004)^{\mbox{\tiny high. ord.}}$~GeV, where the parametric errors, being mainly from experimental origins, 
were summed quadratically. The slight tension with the experimental value may of course be mended by a smaller 
choice of $M_H^{SM}$, although, for $M_H^{SM}\geq114$~GeV, the central value always remains smaller than the 
experimental one.

We will also consider leptonic decay modes of the $Z$ boson, in particular, possible deviations from 
universality between the channels $Z\to e^+e^-$ and $Z\to\tau^+\tau^-$ (which, to our knowledge, has been 
little exploited so far, even in the MSSM). The total width of the $Z$ boson is given experimentally by 
$\Gamma_Z^{\mbox{\tiny exp.}}=(2.4952\pm0.0023)$~GeV. The leptonic decay modes $Z\to l^+l^-$ are characterized 
by the experimental branching ratios \cite{LEP_SLD,PDG}: $BR(Z\to e^+e^-)^{\mbox{\tiny exp.}}=(3.363\pm0.004)
\cdot10^{-2}$ and $BR(Z\to \tau^+\tau^-)^{\mbox{\tiny exp.}}=(3.367\pm0.008)\cdot10^{-2}$. Moreover, the 
relative importance of these modes is determined by $\left(\Gamma(Z\to\tau^+\tau^-)/\Gamma(Z\to e^+e^-)-1
\right)^{\mbox{\tiny exp.}}=(1.9\pm3.2)\cdot10^{-3}$. 

On the theoretical side, the leptonic decay modes are described by the following expression of the decay width 
into a pair of leptons ($l^+l^-$; see {\em e.g.} \cite{Heinemeyer:2004gx}):
\begin{multline}
 \Gamma(Z\to l^+l^-)^{\mbox{\tiny theo.}}=\frac{G_{\mu}(1-\Delta r)}{6\sqrt{2}\pi}M_Z^3\left\{(
g_V^{l,\mbox{\tiny eff}})^2\left[\sqrt{1-\frac{4m_l^2}{M_Z^2}}\left(1+\frac{2m_l^2}{M_Z^2}\right)
+\frac{3}{4\pi}\bar{\alpha}(M_Z)\right]\right.\\
\left.+(g_A^{l,\mbox{\tiny eff}})^2\left[\sqrt{1-\frac{4m_l^2}{M_Z^2}}
\left(1-\frac{4m_l^2}{M_Z^2}\right)+\frac{3}{4\pi}\bar{\alpha}(M_Z)\right]\right\}\label{GamZll}
\end{multline}
$G_{\mu}=(1.16637\pm0.00001)\cdot10^{-5}$ is the Fermi constant measured in muon decays. $\bar{\alpha}(M_Z)\simeq
1/127.92$ is the running fine-structure constant at the $Z$-pole. $\Delta r$ corresponds to the (non-QED) loop 
corrections to this measurement (related to $M_W$: see eq. \ref{mudec} below). $g_V^{l,\mbox{\tiny eff}}$ and 
$g_A^{l,\mbox{\tiny eff}}$ parametrize the $Zl^+l^-$ effective vertex:
\begin{align}
 \Gamma_{Zll}^{\mu}&=\left(\sqrt{2}G_{\mu}(1-\Delta r)\right)^{1/2}M_Z\gamma^{\mu}\left[g_V^{l,\mbox{\tiny eff}}
-\gamma_5g_A^{l,\mbox{\tiny eff}}\right]\label{vertZll}\\
 &=(\sqrt{2}G_{\mu}\rho_l)^{1/2}M_Z\gamma^{\mu}\left[-2Q_l\sin^2\theta^l_{\mbox{\tiny eff}}+I_3^l(1-\gamma_5)
\right]
\end{align}
where we introduce the second parameterization (see {\em e.g.} \cite{EWPO}) in terms of the form factor 
$\rho_l$ and the effective leptonic weak mixing angle $\theta^l_{\mbox{\tiny eff}}$ ($Q_l=-1$ and $I_3^l=-1/2$ 
are respectively the charge, in units of $e$, and the weak isospin of the lepton $l$). At tree level, 
$g_V^{l,\mbox{\tiny tree}}=I_3^l-2Q_l\,s^2_W$ and $g_A^{l,\mbox{\tiny tree}}=I_3^l$.

In the SM, the form factor $\rho_l^{SM}\simeq1.005165$ \cite{ZFitter2} is known with a theoretical uncertainty 
of about $2\cdot10^{-5}$ (with respect to higher orders). All uncertainties considered, we will assume an error 
of $2\cdot10^{-4}$ on this quantity. $\sin^2\theta^{l,SM}_{\mbox{\tiny eff}}$, similarly to $M_W$, has been 
computed beyond two-loop accuracy \cite{sWeffSM} and can be evaluated through numerical approximate formulae 
(see {\em e.g.} eq. 5.7-5.9 in \cite{ZFitter2}), with an estimated theoretical error of $4.9\cdot10^{-5}$. 
Using eq. 5.7-5.9 of \cite{ZFitter2} (with a Higgs mass $M_H^{SM}=150$~GeV), we obtain $\sin^2\theta^{l,SM}_{
\mbox{\tiny eff}}=0.23164\pm(0.00015)^{\mbox{\tiny param.}}\pm(0.00005)^{\mbox{\tiny high. ord.}}$. At the 
level of the branching ratios, we get: $BR(Z\to e^+e^-)^{SM}\simeq3.365\cdot10^{-2}(1\pm1\cdot10^{-3})$ and 
$BR(Z\to \tau^+\tau^-)^{SM}\simeq3.358\cdot10^{-2}(1\pm1\cdot10^{-3})$, where all the errors have been added 
in quadrature.

Alternatively to the branching ratios, one may consider directly the effective leptonic Weinberg angle:
\begin{equation}
 \sin^2\theta^l_{\mbox{\tiny eff}}=\frac{1}{4|Q_l|}\left(1-\frac{g_{V}^{l,\mbox{\tiny eff}}}{
g_{A}^{l,\mbox{\tiny eff}}}\right)\label{sinTW}
\end{equation}
This quantity is extracted from experimental measurements: $(\sin^2\theta^l_{\mbox{\tiny eff}})^{
\mbox{\tiny exp.}}=0.23153\pm0.00016$ \cite{LEP_SLD}. The corresponding SM prediction is quoted in the 
paragraph above eq. \ref{sinTW}.

\section{New Physics contributions to the EWPOs in the NMSSM}
In this section, we shall detail the new contributions, beyond the SM, to $M_W$ and the leptonic $Z$-decays in 
the NMSSM that we take into account and discuss how these are incorporated in our implementation of the 
observables.

\subsection{\boldmath$M_W$ \unboldmath in the NMSSM}
The very precise measurement of the Fermi constant $G_{\mu}$ is extracted from the muon decay 
$\mu\to e\nu_{\mu}\bar{\nu}_e$. In the SM and its extensions, such a decay can also be computed in terms of 
gauge couplings, which allows to relate $G_{\mu}$ to other electroweak quantities (see {\em e.g.} 
\cite{Heinemeyer:2006px} for a more detailed discussion):
\begin{equation}
 \frac{G_{\mu}}{\sqrt{2}}=\frac{\pi\alpha}{M_W^2\left(1-\frac{M_W^2}{M_Z^2}\right)}\left(1+\Delta r(M_W,M_Z,
\ldots)\right)\label{mudec}
\end{equation}
where $\alpha\simeq1/137.03599911$ is the fine-structure constant. $\Delta r$ encompasses the (non-QED) loop 
corrections and is model-dependent. Eq. \ref{mudec} can be inverted, leading to an (implicit, since $\Delta r$ 
depends itself on $M_W$) expression of the $W$ mass in terms of $M_Z$, $G_{\mu}$, etc.:
\begin{equation}
 M_W^2=\frac{M_Z^2}{2}\left[1+\sqrt{1-\frac{4\pi\alpha}{\sqrt{2}G_{\mu}M_{Z}^2}\left(1+\Delta r(M_W,M_Z,
\ldots)\right)}\right]\label{M_W}
\end{equation}
Starting from a reference value $M_W^{2\,(0)}$, for which we choose $M_W^{2\,(0)}=M_W^{2\,SM}(M_H^{SM}=
150~\mbox{GeV})$, the contribution to eq. \ref{M_W} at leading-order in $\delta M_W^2=M_W^2-M_W^{2\,(0)}$ 
({\em i.e.} in $\Delta r$) can be obtained directly by evaluating $\Delta r(M_W^{2\,(0)})$. Only at higher 
order need we take the shift of the right-hand-side of eq. \ref{M_W}, induced by $\delta M_W^2$, into account. 
The traditional approach consists then in iterating eq. \ref{M_W} a few times. An alternative method consists 
in Taylor-expanding the right-hand-side of eq. \ref{M_W} at $M_W^2=M_W^{2\,(0)}+\delta M_W^2$: for a 
calculation of eq. \ref{M_W} at second order in $\delta M_W$, $\Delta r$ needs to be expanded at first order, 
leading to terms $\propto\Delta r(M_W^{2\,(0)})$, which should consistently be evaluated at second order, and 
terms $\propto d\Delta r/dM_W^2\,(M_W^{2\,(0)})$, which need only be evaluated at leading order. We compared 
both methods, which, given the uncertainty, led to similar results.

Given the far-superior (beyond two-loop) accuracy of the SM calculation, we split $\Delta r^{NMSSM}=
\Delta r^{SM}+\Delta r^{NP}$, where $\Delta r^{NP}$ contains all the new-physics contributions (and substracts 
the absent SM contributions, in particular from the SM Higgs). At leading (one-loop) order,
\begin{equation}
 \Delta r^{NP,1l.}=\left[\frac{\hat{\Sigma}_W(0)}{M_W^2}+2\delta^{v}+\delta^{b}\right]_{\mbox{\tiny NMSSM-SM}}
\end{equation}
where $\hat{\Sigma}_W(0)$ is the (transverse) W renormalized self-energy at $0$ momentum while $\delta^{v}$ and 
$\delta^{b}$ summarize vertex and box corrections to the decay $\mu\to e\nu_{\mu}\bar{\nu}_e$ (here, we will 
neglect the light lepton masses and assume universality in the slepton spectrum for the first two generations). 
$\Delta r^{NP,1l.}$ may be further separated into contributions from the modified Higgs sector (the 
corresponding contributions to $\delta^{v}$ and $\delta^{b}$ vanish in the approximation $m_e=0=m_{\mu}$) and 
contributions from the supersymmetric particles, sleptons, charginos and neutralinos. We computed these 
one-loop contributions anew and found agreement with the existing literature (see {\em e.g.} \cite{Cao:2008rc}). 
The corresponding calculation is formally identical to that in the MSSM \cite{MWMSSM_1L}, were it not for the 
presence of two additional Higgs and one additional neutralino states in the NMSSM. Note that in the MSSM, 
non-minimal flavour-violating \cite{Heinemeyer:2004by} and CP-violating \cite{Kang:2000bt,Heinemeyer:2006px} 
effects were also investigated.

Higher order corrections should be included through an expansion of the resummation formula \cite{Consoli:1989fg}: 
$ 1+\Delta r=\left((1+\Delta\alpha)(1+\frac{c_W^2}{s_W^2}\Delta\rho)-\Delta r_{\mbox{\tiny rem.}}\right)^{-1}$. 
Corrections of order $\alpha\alpha_S$ to the parameter $\Delta\rho$ were computed for the MSSM 
\cite{DrhoMSSM_2LQCD}: beyond the SM-like effects, they arise from gluon/squark and gluino/squark diagrams. 
Such contributions are identical in the NMSSM and the MSSM, hence may be included in the calculation directly. 
However, we will always consider heavy gluinos so that the gluino effects remain negligible. Additional 
$O(Y_iY_j/(4\pi)^2)$ contributions, $Y_{i,j}=Y_t,Y_b$ being the top and bottom Yukawa couplings, were 
calculated in the MSSM \cite{DrhoMSSM_2LHQ}, including SM-like quark/Higgs as well as supersymmetric 
higgsino/squark diagrams. Such effects cannot be incorporated so straightforwardly in the NMSSM. However, we 
note that most of such effects are captured by the corresponding SM contribution, contained in $\Delta r^{SM}$,
that we will not substract.

To compute the error estimate, we considered the analysis in \cite{Heinemeyer:2007bw} for the MSSM as well as 
the magnitude, for the MSSM, of those higher-order contributions which are known in the MSSM but neglected in 
our computation. We therefore decided to include a higher-order uncertainty of about $10$~{MeV} (when squarks 
are all heavier than $\sim500$~GeV) to $20$~MeV (for squark masses below $\sim500$~GeV) for $M_W$. To define 
our ``exclusion'' limits, we double the SM parametric error and add linearly the SM and SUSY higher-order 
uncertainties.

\subsection{New Physics contributions to \boldmath$g_{V,A}^{l,\mbox{\tiny eff}}$\unboldmath}
We already introduced the general formalism associated with $Z\to l^+l^-$ decays (eq. \ref{GamZll},
\ref{vertZll}). Again, we split $g_{V,A}^{l,\mbox{\tiny eff}}=(g_{V,A}^{l,\mbox{\tiny eff}})^{SM}+
(g_{V,A}^{l,\mbox{\tiny eff}})^{NP}$. Note however that a ``hidden contribution'' already intervenes at 
tree-level in $g_{V}^{l,\mbox{\tiny eff}}$: as we analysed in the previous subsection, radiative corrections 
induce a shift in the determination of $M_W$, which, in turn, results in a shift of the Weinberg angle.

The new physics contributions to $g_{V,A}^{l,\mbox{\tiny eff}}$ were computed (at the $Z$-pole) at the one-loop 
level according to (see {\em e.g.} \cite{Cao:2008rc}):
\begin{equation}
\begin{cases}
 (g_{V}^{l,\mbox{\tiny eff}})^{NP}=\left[\Gamma_V(M_Z^2)-g_V^{l,\mbox{\tiny tree}}\Sigma_V^l(m_l^2)-
g_A^{l,\mbox{\tiny tree}}\Sigma_A^l(m_l^2)-g_A^{l,\mbox{\tiny tree}}\frac{c_W}{s_W}
\frac{\Sigma^{\gamma Z}(0)}{M_Z^2}-\frac{1}{2}g_V^{l,\mbox{\tiny tree}}\hat{\Sigma}'_Z(M_Z^2)\right.\\
\hspace{10.7cm}\left.-2\frac{c_Ws_W}{M_Z^2}\hat{\Sigma}^{\gamma Z}(M_Z^2)\right]_{NMSSM-SM}\\
 (g_{A}^{l,\mbox{\tiny eff}})^{NP}=\left[\Gamma_A(M_Z^2)-g_V^{l,\mbox{\tiny tree}}\Sigma_A^l(m_l^2)-
g_A^{l,\mbox{\tiny tree}}\Sigma_V^l(m_l^2)-g_A^{l,\mbox{\tiny tree}}\frac{c_W}{s_W}
\frac{\Sigma^{\gamma Z}(0)}{M_Z^2}-\frac{1}{2}g_V^{l,\mbox{\tiny tree}}\hat{\Sigma}'_Z(M_Z^2)
\right]_{NMSSM-SM}
\end{cases}
\end{equation}
where $\Gamma_{V,A}(M_Z^2)$ denote the $Zll$ (1PI) vertex corrections at the $Z$-pole, $\Sigma_{V,A}^l$ 
stand for the lepton ($l$) self-energies, $\Sigma_Z$ and $\Sigma^{\gamma Z}$, for the $Z$ and $\gamma Z$ 
self-energies. The hat $\hat{}$ signals renormalized self-energies and the prime $'$, derivatives with respect 
to the external 4-momentum squared. The indices $V$ and $A$ still denote vector and axial(-vector) contributions.

Again, the new contributions arise from the NMSSM Higgs sector (from which the 1-loop contribution of the SM 
Higgs is substracted) as well as supersymmetric particles ({\em e.g.} Higgs/lepton and chargino or 
neutralino/sleptons diagrams in $\Gamma_{V,A}$). Our calculation agrees with the existing literature (see 
{\em e.g.} \cite{Cao:2008rc}). However, we also incorporate higher-order corrections in the form of corrected 
Yukawa couplings, relevant for large $\tan\beta$ (see {\em e.g.} \cite{Marchetti:2008hw}):
\begin{equation}
 Y_l=\frac{m_l\tan\beta}{\mbox{\em v}\left(1+\varepsilon_l\tan\beta\right)}
\end{equation}

We allow for a $30\%$ uncertainty on the gauge contributions to $(g_{V,A}^{l,\mbox{\tiny eff}})^{NP}$, where 
QCD corrections may intervene at the two-loop level. However, we confine ourselves to a $10\%$ error on the 
leptonic part ($\Gamma_{V,A}$ and $\Sigma_{V,A}^l$) where no strong coupling should arise at the following 
order. Error estimates on $(g_{V,A}^{l,\mbox{\tiny eff}})^{SM}$ are extracted from the uncertainty on 
$\rho_l^{SM}$ and $\sin^2\theta^{l,SM}_{\mbox{\tiny eff}}$, which was presented in the introduction. These SM 
and NP uncertainties are added linearly. At the level of the branching ratios, they are then combined 
quadratically with the ($2\,\sigma$) error on the experimental input ($M_Z$, $G_{\mu}$, $\Gamma_Z$) to define 
the error estimate.

The observable $\left(\Gamma(Z\to\tau^+\tau^-)/\Gamma(Z\to e^+e^-)-1\right)$ should be sensitive to possible 
deviations from universality. Using eq. \ref{GamZll} (at first order in $m_{\tau}^2/M_Z^2$ and neglecting 
$m_e^2\ll m_{\tau}^2,M_Z^2$), we derive:
\begin{equation}
\displaystyle \frac{\Gamma(Z\to\tau^+\tau^-)}{\Gamma(Z\to e^+e^-)}-1=\frac{\left[(g_{V}^{\tau,\mbox{\tiny eff}}
)^2-(g_{V}^{e,\mbox{\tiny eff}})^2+(g_{A}^{\tau,\mbox{\tiny eff}})^2-(g_{A}^{e,\mbox{\tiny eff}})^2\right]
-\frac{6m_{\tau}^2(g_{A}^{\tau,\mbox{\tiny eff}})^2}{M_Z^2\left(1+\frac{3}{4\pi}\bar{\alpha}(M_Z)\right)}
}{(g_V^{e,\mbox{\tiny eff}})^2+(g_A^{e,\mbox{\tiny eff}})^2}
\end{equation}
To estimate the error on $(g_{V,A}^{\tau,\mbox{\tiny eff}})^2-(g_{V,A}^{e,\mbox{\tiny eff}})^2\simeq2\,
g_{V,A}^{e,\mbox{\tiny eff}}\left(g_{V,A}^{\tau,\mbox{\tiny eff}}-g_{V,A}^{e,\mbox{\tiny eff}}\right)$, we 
retain $10\%$ of the universality-deviating contributions (Higgs and supersymmetric sectors, with no QCD 
corrections at two-loop) in $g_{V,A}^{\tau,\mbox{\tiny eff}}-g_{V,A}^{e,\mbox{\tiny eff}}$, to which we add a 
fixed error amounting to $10^{-3}\cdot g_{V,A}^{e,\mbox{\tiny eff}}$ for neglected orders of a different type.

\section{Numerical results}
The observables are included in a fortran code as prescribed in the previous section. This program is then 
coupled to the NMSSMTools 2.3.2 package \cite{NMSSMTools}, computing the NMSSM spectrum and checking already 
several phenomenological constraints: stability of the electroweak vacuum, bounds from LEP \cite{LEPHMSSM}, and 
in particular the new {\sc Aleph} limits \cite{ALEPH:2010aw}\footnote{A bug in the corresponding version of the 
code for the computation of {\sc Aleph} constraints was replaced by a self-made program. This problem was fixed, 
in the meanwhile, in the version 2.3.3 of NMSSMTools.}, from $B$- and $\Upsilon$-physics \cite{Bphys,Upsilon}, 
etc. Note that the NMSSM contribution to the anomalous magnetic moment of the muon is also computed 
\cite{Domingo:2008bb}. However, given the still unclear situation concerning the agreement of $\tau$- and 
$e^+e^-$-based SM hadronic contributions to this observable, we will regard a SM-like result as allowed 
(although $\gsim3\,\sigma$ away from the experimental measurement, when using $e^+e^-$-data; see {\em e.g.} 
\cite{Davier:2010nc} for a recent analysis). First, we will discuss $M_W$ and the leptonic $Z$-decays in the 
MSSM limit of the NMSSM. Then, we will turn to specific NMSSM effects.

\subsection{MSSM limit}
The MSSM limit of the NMSSM is obtained by choosing small parameters $\lambda\sim\kappa\ll1$ while keeping 
$\mu_{eff}\propto\lambda/\kappa$ at the electroweak-TeV scale. In this case, the NMSSM would be 
undistinguishable, in collider phenomenology, from a genuine MSSM, except for possible dark matters effects
(in case of a singlino LSP). Note that most NMSSM effects in the EWPOs ({\em e.g.} from supersymmetric 
particles) may be expected to be MSSM-like and this subsection is meant to summarize such results as well as 
serve for comparison with specific NMSSM effects, which will be discussed separately.

\begin{figure}[t]
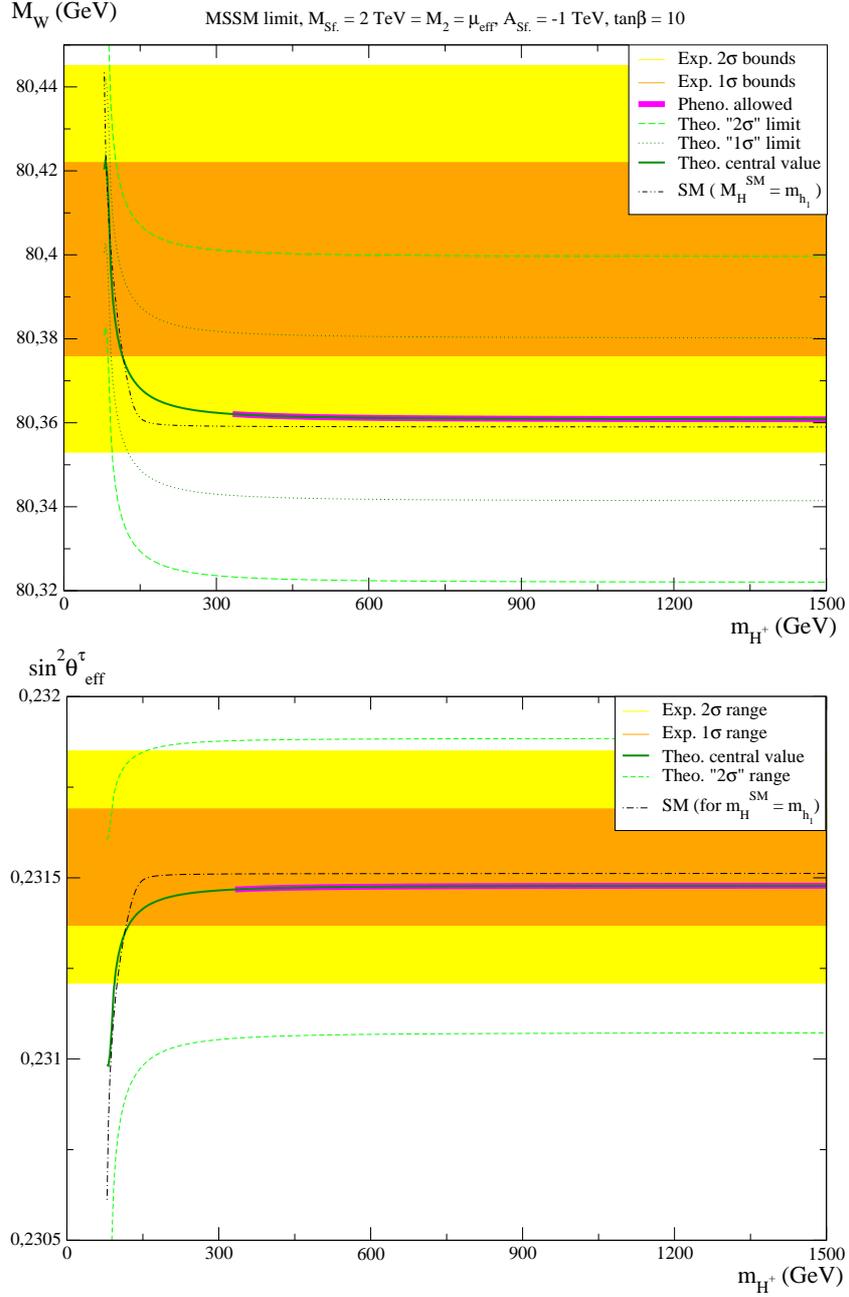

 \begin{center}
\includegraphics[width=11cm]{MW_mH+.eps}
\includegraphics[width=11cm]{s2tweff_mH+.eps}
 \end{center}
\caption{Impact of the Higgs sector on $M_W$ and $\sin^2\theta^{\tau}_{\mbox{\tiny eff}}$, in the MSSM limit. 
Supersymmetric effects are suppressed by the choice of a heavy SUSY scale $\sim2$~TeV. Here and below, the 
experimental $1\,\sigma$ and $2\,\sigma$ ranges are depicted in orange and yellow, respectively. The dark-green 
curve corresponds to the theoretical prediction, flanked by its light-green dashed uncertainty limits (the 
dotted dark-green curves in the case of the $M_W$ plot correspond to a ``$1\,\sigma$'' half-error). The SM 
contribution, for a SM Higgs mass equal to the lightest MSSM-like CP-even Higgs state ($M_H^{\mbox{\tiny SM}}=
m_{h_1^0}$; note that, on the ``plateau'' when $m_{H^{\pm}}\geq250$~GeV, $m_{h_1^0}\sim117$~GeV) is represented 
by the dot-dash black curves.  The ``pink aura'' around the dark-green curve corresponds to points which were 
found in agreement with all constraints implemented in NMSSMTools: the lower mass-range is excluded by LEP 
constraints on Higgs searches.}
\label{fig:obs(mH+)}
\end{figure}

The impact of a MSSM-like Higgs sector is illustrated by Fig. \ref{fig:obs(mH+)}: $M_W$ and 
$\sin^2\theta^{\tau}_{\mbox{\tiny eff}}$ are plotted against the charged-Higgs mass $m_{H^{\pm}}$ (controlling 
the mass of the heavy MSSM-like approximate $SU(2)_L$ doublet). The supersymmetric spectrum was chosen very 
heavy with sfermion, chargino and neutralino masses of about $\sim2$~TeV, so as to suppress associated effects. 
Consequently the effects are found to be very small as long as $m_{H^{\pm}}\gsim250$~GeV, with {\em e.g.} $M_W$ 
almost entirely determined by the SM-result applied to its lightest doublet mass ($\sim117$~GeV in the 
considered case). Low values of $m_{H^{\pm}}$ (as the whole Higgs spectrum becomes light) would tend to 
increase $M_W$ (hence improving its agreement with the experimental value), decrease $\sin^2\theta^{\tau}_{
\mbox{\tiny eff}}$ (hence worsening its agreement with experiment), but such effects become significant mainly 
for mass ranges already excluded by LEP constraints. We therefore conclude that the impact of the MSSM Higgs 
sector is mainly driven by the lightest CP-even (doublet) Higgs, whose mass is almost as strongly bound by 
LEP-constraints as that of its SM alter ego. In contrast, unconventional scenarii are possible in the NMSSM 
CP-even Higgs spectrum, and this is where specific NMSSM effects are expected.

\begin{figure}[t]
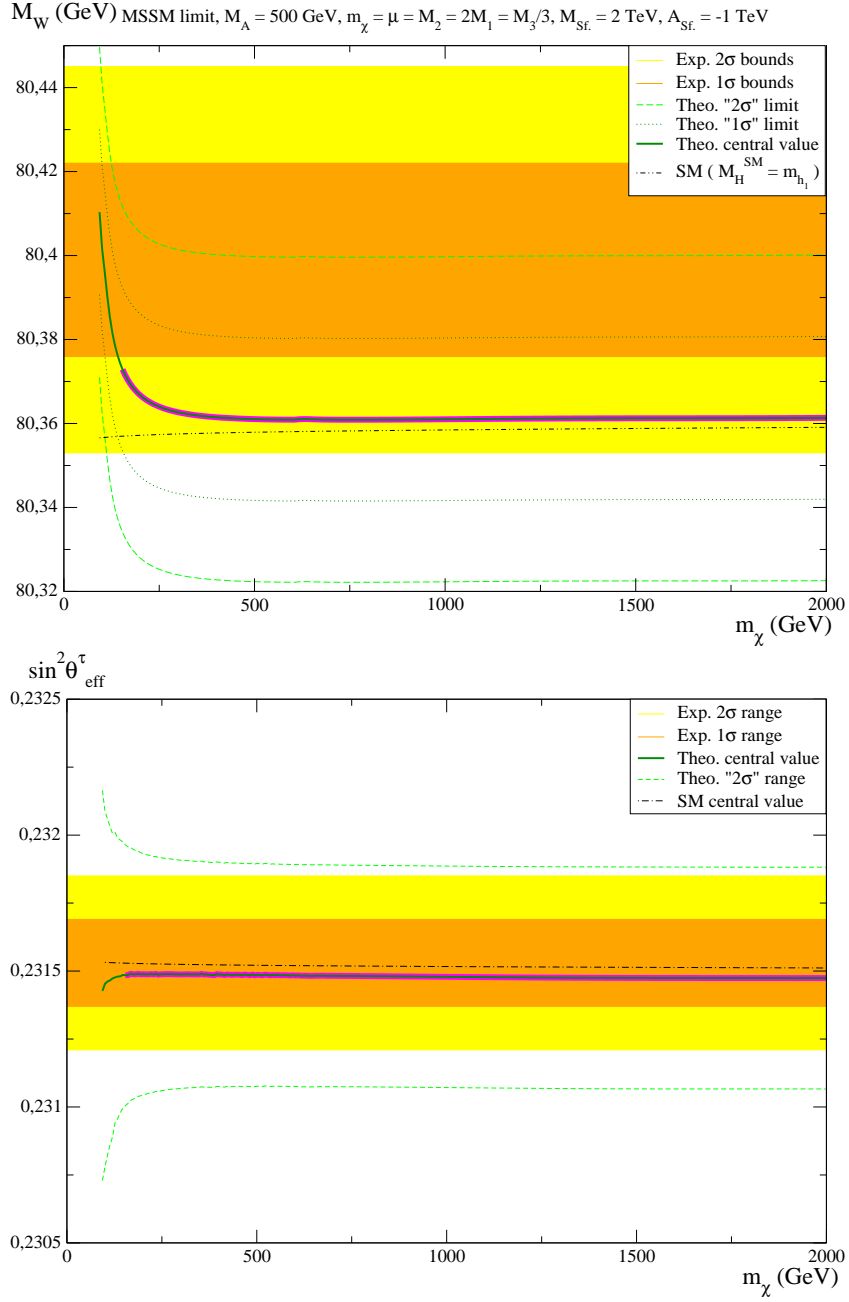

 \begin{center}
\includegraphics[width=11cm]{MW_mchi.eps}
\includegraphics[width=11cm]{s2tweff_mchi.eps}
 \end{center}
\caption{Impact of the gaugino-higgsino sector on $M_W$ and $\sin^2\theta^{\tau}_{\mbox{\tiny eff}}$ in the 
MSSM limit. A common scale $m_{\chi}\equiv\mu_{\mbox{\tiny eff}}=M_2=2M_1=M_3/3$ was used for the scan. Here 
and below, the heavy Higgs doublet is set at a scale $\sim500$~GeV. The Sfermion sector is still very heavy 
($\sim2$~TeV). The colour-code is similar to that of Fig. \ref{fig:obs(mH+)}. The lower mass range is excluded 
by the experimental lower-bounds on chargino/neutralino masses.}
\label{fig:obs(mchi)}
\end{figure}

\begin{figure}[t]
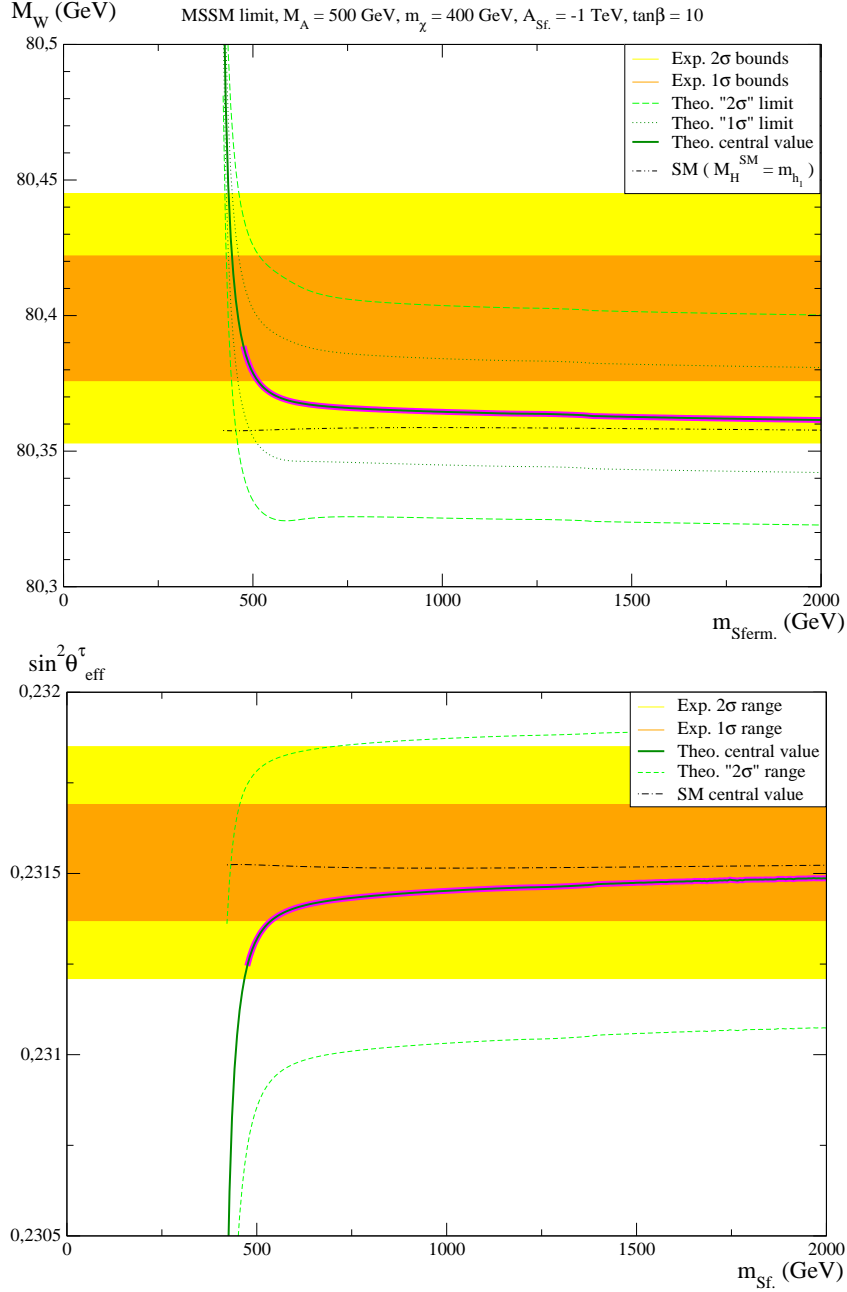

 \begin{center}
\includegraphics[width=11cm]{MW_mSf.eps}
\includegraphics[width=11cm]{s2tweff_mSf.eps}
 \end{center}
\caption{Impact of the sfermions on $M_W$, $\sin^2\theta^{\tau}_{\mbox{\tiny eff}}$ in the MSSM limit. The scan 
was performed over a universal soft SUSY-breaking sfermion mass $m_{\mbox{\tiny Sf.}}$, the trilinear couplings 
remaining at $-1$~TeV (Therefore, the stop sector always provides the lightest and the heaviest sfermion. Note 
that the sfermion contribution to the lightest Higgs mass remains sufficient to circumvent LEP bounds, 
essentially due to heavy states, with mass over $500$~GeV). $m_{\chi}\equiv\mu_{\mbox{\tiny eff}}=M_2=2M_1=M_3
/3=400$~GeV. The colour-code remains as in Fig. \ref{fig:obs(mH+)}. Constraints on the masses of sfermion 
states exclude the lower mass-range.}
\label{fig:obs(mSf)}
\end{figure}

We then consider effects associated with the supersymmetric spectrum (and settle for a heavy Higgs doublet 
mass of $\sim500$~GeV): such contributions are essentially common to the MSSM and the NMSSM, since the 
NMSSM spectrum, except for an additional (but usually almost decoupled) neutralino state (the singlino), is 
similar to that of the MSSM. In Fig. \ref{fig:obs(mchi)}, the sfermion sector is kept very heavy $\sim2$~TeV 
and we vary the mass of the supersymmetric fermions (gauginos/higgsinos) according to the hierarchy 
$2M_1=M_2=\mu_{\mbox{\tiny eff}}=M_3/3\equiv m_{\chi}$. We find that a light gauginos-higgsino sector can 
increase $M_W$ of a few $10$~MeV, while its effect on the other observables remains small (note that this will 
not remain the case for lighter sfermion spectra). In Fig. \ref{fig:obs(mSf)}, we scan over a common soft 
SUSY-breaking sfermion mass $m_{\mbox{\tiny Sf.}}$. Note, however, that the left-right sfermion mixing may 
remain large (trilinear soft couplings are kept at $-1$~TeV, while $\mu_{\mbox{\tiny eff}}=400$~GeV), so that 
even for small $m_{\mbox{\tiny Sf.}}$, one of the stops (and sbottoms) is sensibly heavy, sufficiently to generate a CP-even 
Higgs mass beyond LEP bounds. Lighter sfermion spectra contribute to the predicted $M_W$, and may improve its 
agreement with the experimental measurement, but they correspondingly increase the tension in $\sin^2\theta^{
\tau}_{\mbox{\tiny eff}}$, which (expectedly) shows a behaviour ``opposite'' to $M_W$. Note that, as $M_W$ 
increases, the $SU(2)_L$ interaction strengthens, leading to a larger $BR(Z\to\tau^+\tau^-)$ (which is 
not shown but essentially remains within the $1\,\sigma$ experimental range). No deviation from lepton 
universality is observed: note that slepton states are still heavy while the lightest stop mass approaches the 
electroweak scale (due to the more efficient stop mixing). We stress that these supersymmetric effects 
generically prove the most significant contributions in both the MSSM and the NMSSM, and explain how 
supersymmetric extensions of the SM, albeit often subject to comparable limits on the mass of the lightest 
CP-even Higgs as the SM, can improve the agreement of the theoretical predictions for EWPOs (and more 
specifically $M_W$) with the corresponding experimental values.

\begin{figure}[t]
 \begin{center}
\includegraphics[width=11cm]{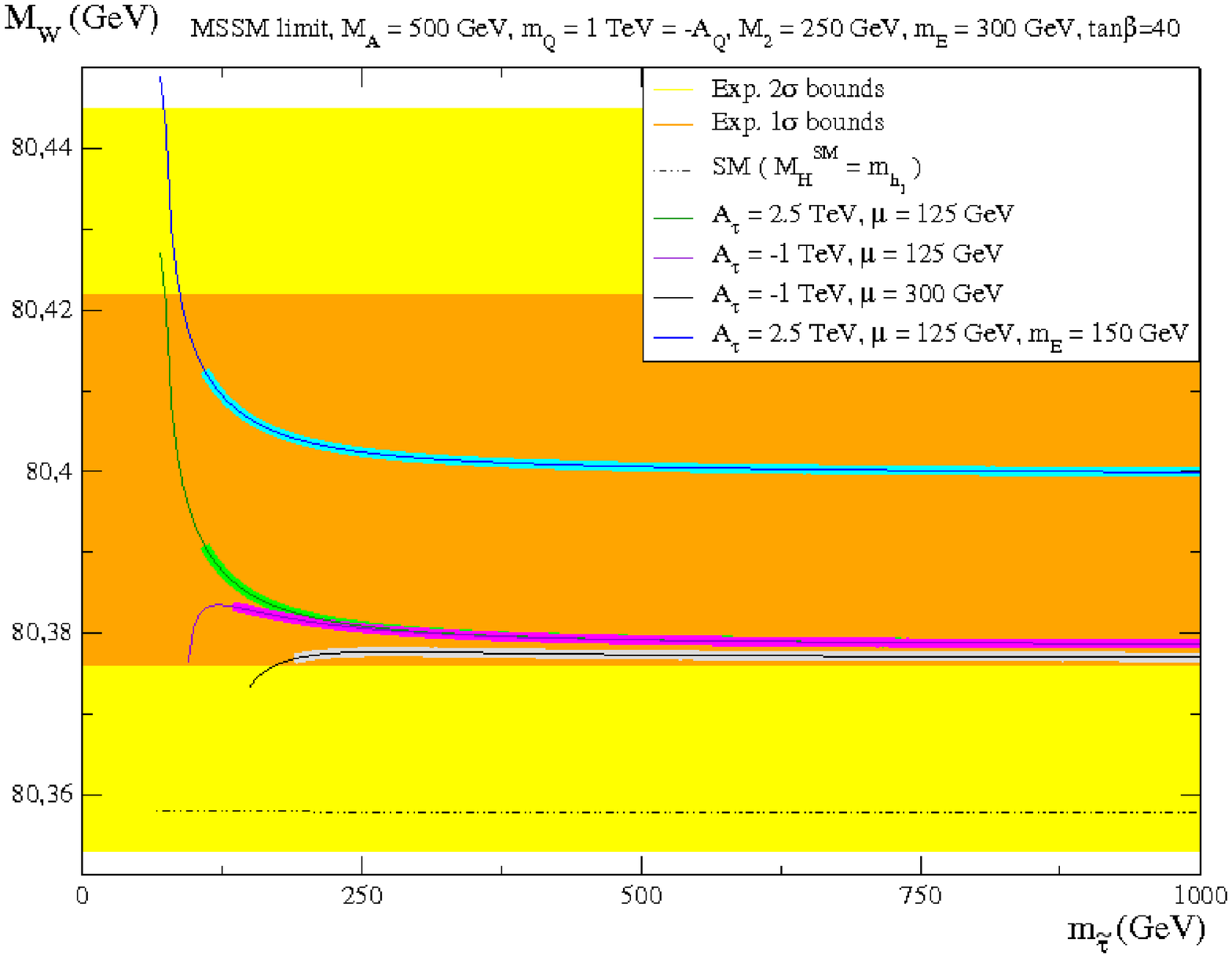}
\includegraphics[width=11cm]{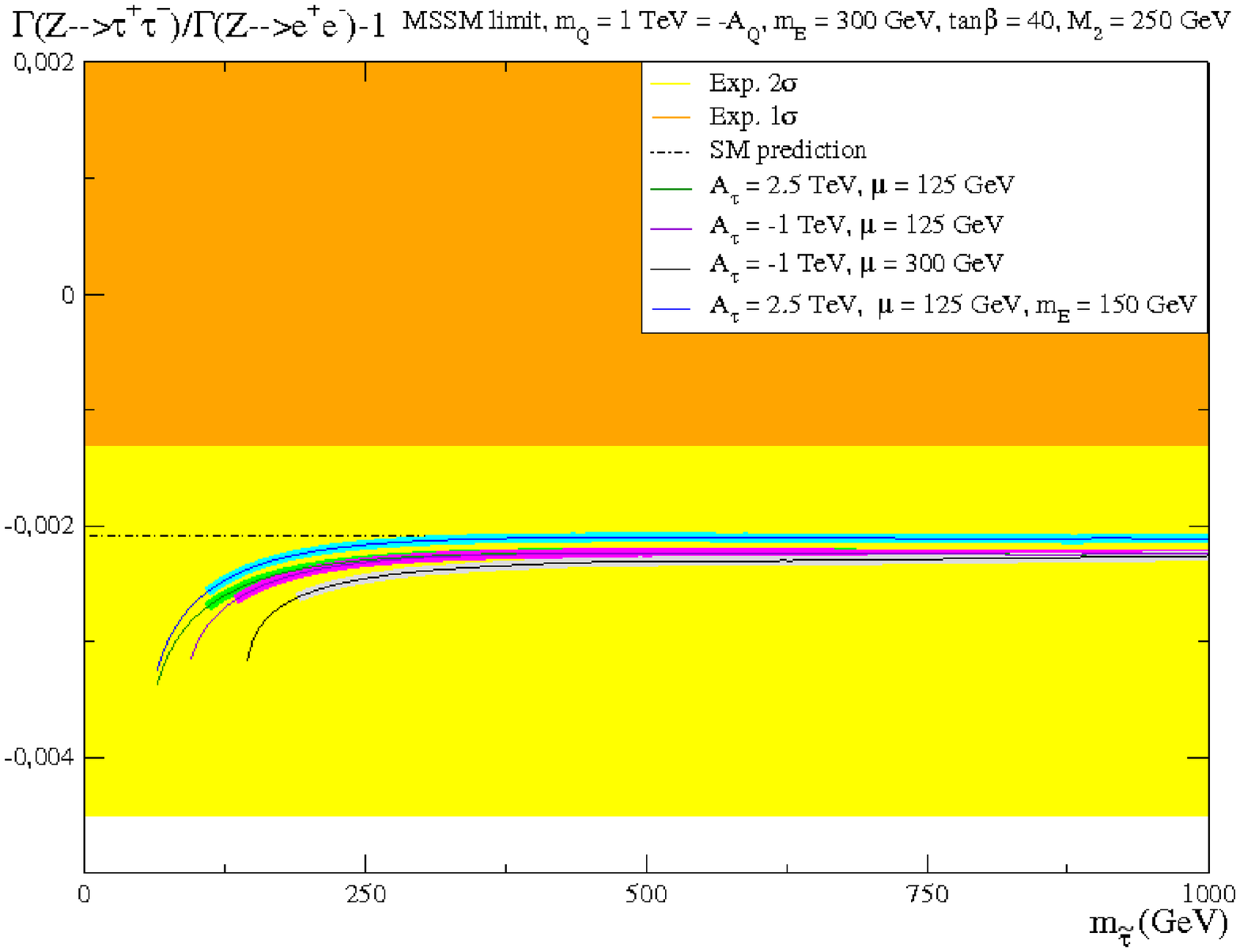}
 \end{center}
\caption{Impact of the slepton sector on $M_W$ and $\Gamma(Z\to\tau^+\tau^-)/\Gamma(Z\to e^+e^-)-1$ in the MSSM 
limit. The scan was performed over a soft SUSY-breaking (left- and right-) stau mass $m_{\tilde{\tau}}$, for 
several values of $\mu_{\mbox{\tiny eff}}$ and the trilinear coupling $A_{\tau}$, in order to vary the stau 
mixing: $(A_{\tau},\mu_{\mbox{\tiny eff}})$ is chosen as $(2.5~\mbox{TeV},125~\mbox{GeV})$ for the green curve 
(small mixing), $(-1~\mbox{TeV},125~\mbox{GeV})$ for the violet curve, $(-1~\mbox{TeV},300~\mbox{GeV})$ for the 
black curve (large mixing). The sleptons of the first two generations have a soft mass of $m_{\tilde{E}}=
300$~GeV. The blue curve is similar to the green one, but for $m_{\tilde{E}}=150$~GeV. As before, the ``auras'' 
alongside the curves correspond to points passing the constraints of NMSSMTools (the lower mass-range being 
excluded by bounds on slepton masses). The theoretical error bars are not displayed, but the uncertainty is 
similar in magnitude to that of {\em e.g.} Fig \ref{fig:obs(mH+)}.}
\label{fig:obs(mStau)}
\end{figure}

Now, we wish to consider flavour effects and, since we consider leptonic observables, we vary the common soft 
SUSY-breaking stau mass $m_{\tilde{\tau}}$ and show the impact on $M_W$ and $\Gamma(Z\to\tau^+\tau^-)
/\Gamma(Z\to e^+e^-)-1$ in Fig. \ref{fig:obs(mStau)}. Several choices of the trilinear $A_{\tau}$ coupling and 
$\mu_{\mbox{\tiny eff}}$, regulating the mixing of staus ($\propto m_{\tau}\left(A_{\tau}-
\mu_{\mbox{\tiny eff}}\tan\beta\right)$), are studied. We choose a large value for $\tan\beta=40$, enhancing 
leptonic Yukawa couplings. Deviations of $M_W$ occur for small stau masses and can reach up to $\sim10$~MeV in 
the allowed slepton mass range. We observe that this effect can be suppressed, even reversed, when mixing among 
the staus becomes efficient. We also notice that a light slepton ($\sim300$~GeV) sector is sufficient to bring 
the theoretical prediction of $M_W$ within its experimental $1\,\sigma$ error bars (at the expense of $\sin^2
\theta^{\tau}_{\mbox{\tiny eff}}$ however, which is not shown but may lead to difficulties for very light 
sleptons $\sim150$~GeV). Deviations from lepton universality in $Z$-decays are generated for light slepton 
spectra and worsen the SM results with respect to the experimental limits. However, such effects remain 
essentially small and, given the current uncertainties, are not significant.

\begin{figure}[t]
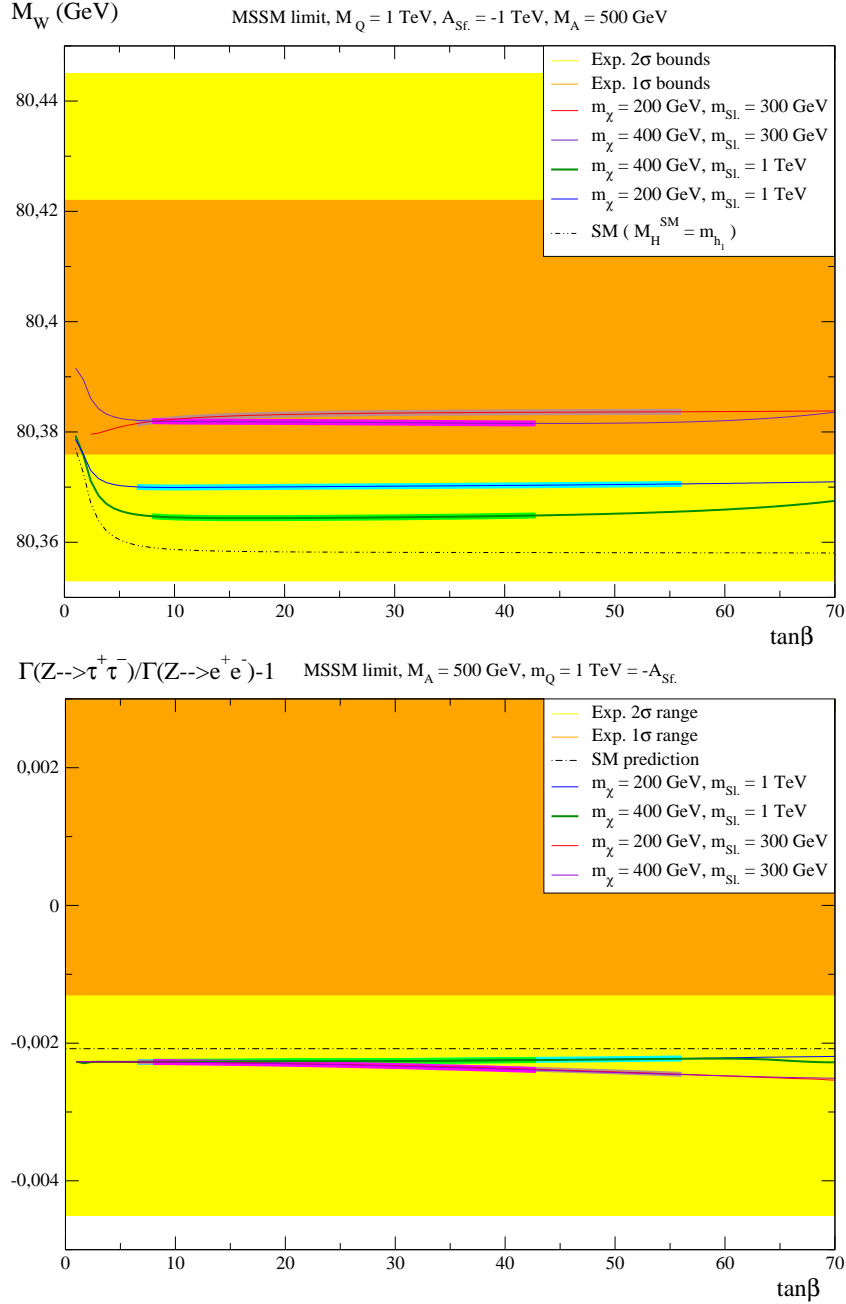

 \begin{center}
\includegraphics[width=11cm]{MW_tanB.eps}
\includegraphics[width=11cm]{ratio_tanB.eps}
 \end{center}
\caption{Impact of $\tan\beta$ on $M_W$ and $\Gamma(Z\to\tau^+\tau^-)/\Gamma(Z\to e^+e^-)-1$ in the MSSM limit. 
The scan was performed for several slepton and chargino/neutralino scales: $(m_{\chi},m_{\tilde{E},
\tilde{\tau}})=(400~\mbox{GeV},1~\mbox{TeV})$ for the green curve, $(200~\mbox{GeV},1~\mbox{TeV})$, 
$(400~\mbox{GeV},300~\mbox{GeV})$ and $(200~\mbox{GeV},300~\mbox{GeV})$ respectively for the blue, violet and 
red ones. Constraints applying at low $\tan\beta$ originate from bounds on the lightest Higgs mass, while the 
large $\tan\beta$ regime is excluded by $B$-physics processes ($BR(\bar{B}^0_s\to\mu^+\mu^-)$).}
\label{fig:obs(tanB)}
\end{figure}

Finally, Fig. \ref{fig:obs(tanB)} shows the variations of $M_W$ and $\Gamma(Z\to\tau^+\tau^-)/\Gamma(Z\to e^+
e^-)-1$ in terms of $\tan\beta$ for heavier/lighter slepton and gaugino-higgsino sectors. The significant 
effects (in $M_W$) observed at low $\tan\beta$ correspond to light CP-even Higgs masses, already excluded by 
LEP. Light supersymmetric spectra tend to shift $M_W$ upwards as we already discussed. Small flavour effects 
appear at large $\tan\beta$ (in $\Gamma(Z\to\tau^+\tau^-)/\Gamma(Z\to e^+e^-)-1$) for light sleptons, but 
remain negligible in comparison to the uncertainty, all the more that large $\tan\beta$ regions are already 
constrained by $B$-physics observables (such as $BR(\bar{B}^0_s\to\mu^+\mu^-)$).


As a summary for this analysis of the MSSM limit (hence of MSSM-like NMSSM effects), most of the MSSM, within 
its allowed parameter space (in view of bounds on direct searches of Higgs and supersymmetric particles, 
$B$-physics), lies within the allowed range of the considered observables, and only weak constraints can thus 
be extracted, except perhaps for very light sfermion spectra (in view of $\sin^2\theta^{\tau}_{\mbox{\tiny eff}
}$). Future experimental developments may lead to more significant limits, although a comparable accuracy from 
the theoretical side might prove difficult to achieve (note, however, that a large portion of the theoretical 
error is parametric and will therefore be reduced along with the experimental uncertainty on $m_t$, etc.). New 
physics effects lie essentially in light supersymmetric spectra ($\sim100-300$~GeV) and may improve the 
agreement of the theoretically predicted $M_W$ with its experimental determination. Note however that tensions 
might then correspondingly arise in $\sin^2\theta^{\tau}_{\mbox{\tiny eff}}$. Higgs effects are mainly 
controlled by the mass of the SM-like CP-even Higgs. Finally, effects violating lepton universality tend to 
increase the tension with experimental measurements, but remain small (below the uncertainty) in the available 
parameter space: universal contributions are found to dominate.

\subsection{NMSSM at low \boldmath$\tan\beta$\unboldmath}\label{subsec:largelamb}
We will now consider specific NMSSM effects, hence use $\lambda$ (or $\kappa$) of order $1$ (Note that, in order 
to ensure the perturbativity of such couplings up to a Grand-Unification scale, one must require $\lambda\lsim
0.8$: see {\em e.g.} Fig. 1 of \cite{Ellwanger:2009dp}). These new effects are essentially expected in the 
Higgs sector since, except for the additional singlino state, the rest of the spectrum is essentially MSSM-like. 
Note however, that a first consequence of the modified Higgs sector consists in an enlarged parameter space, 
allowing for low values of $\tan\beta\sim1-2$, despite LEP constraints: large $\lambda$ generates indeed an 
additional contribution to the light CP-even doublet Higgs mass at tree-level, so that the maximal mass is 
actually reached at low $\tan\beta$, provided $\lambda\mbox{\em v}\gsim M_Z$. In contrast, the MSSM cannot 
generate sufficiently large Higgs masses in this regime, leading to the exclusion of low $\tan\beta$.
\begin{figure}[t]
 \begin{center}
\includegraphics[width=11cm]{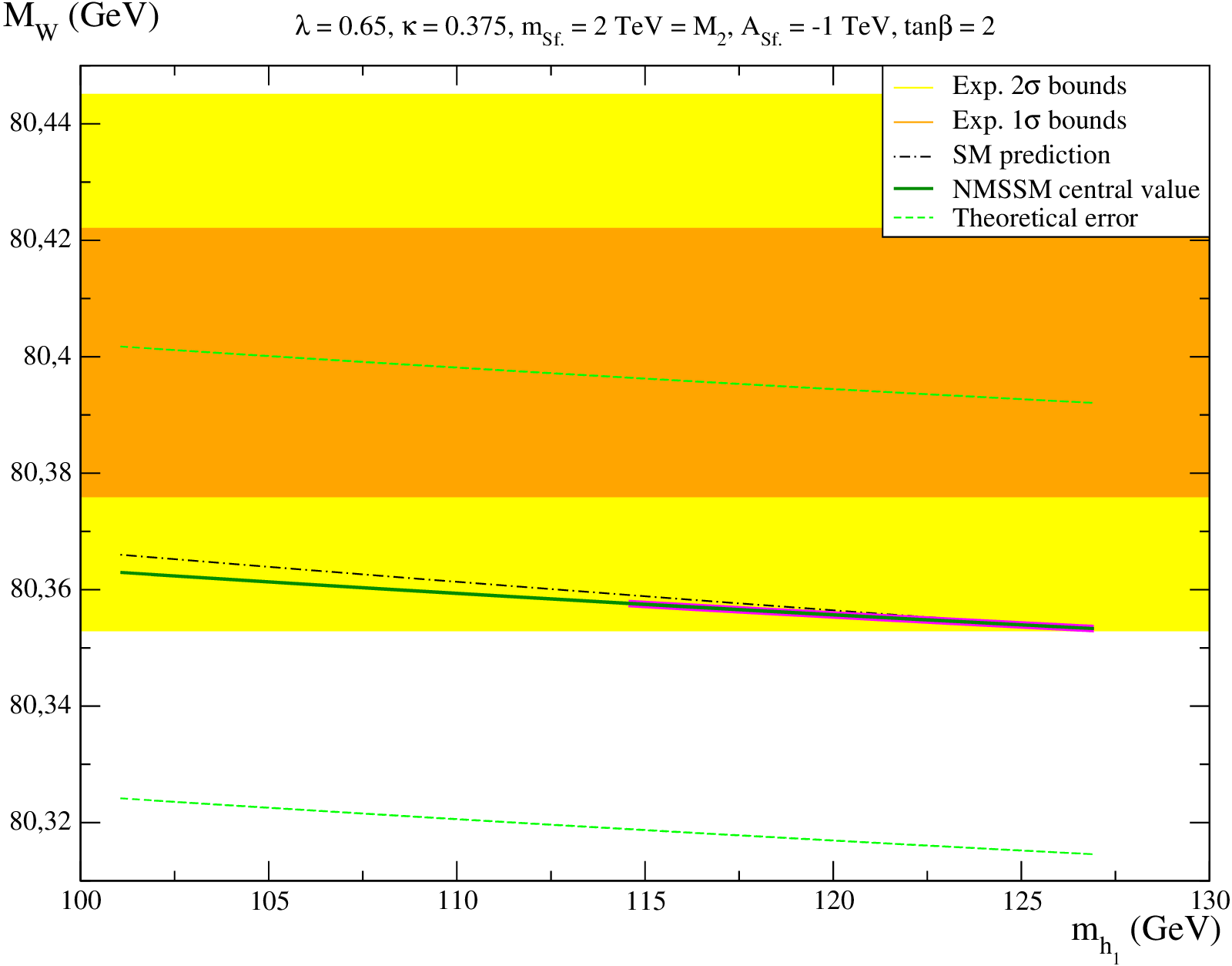}
\includegraphics[width=11cm]{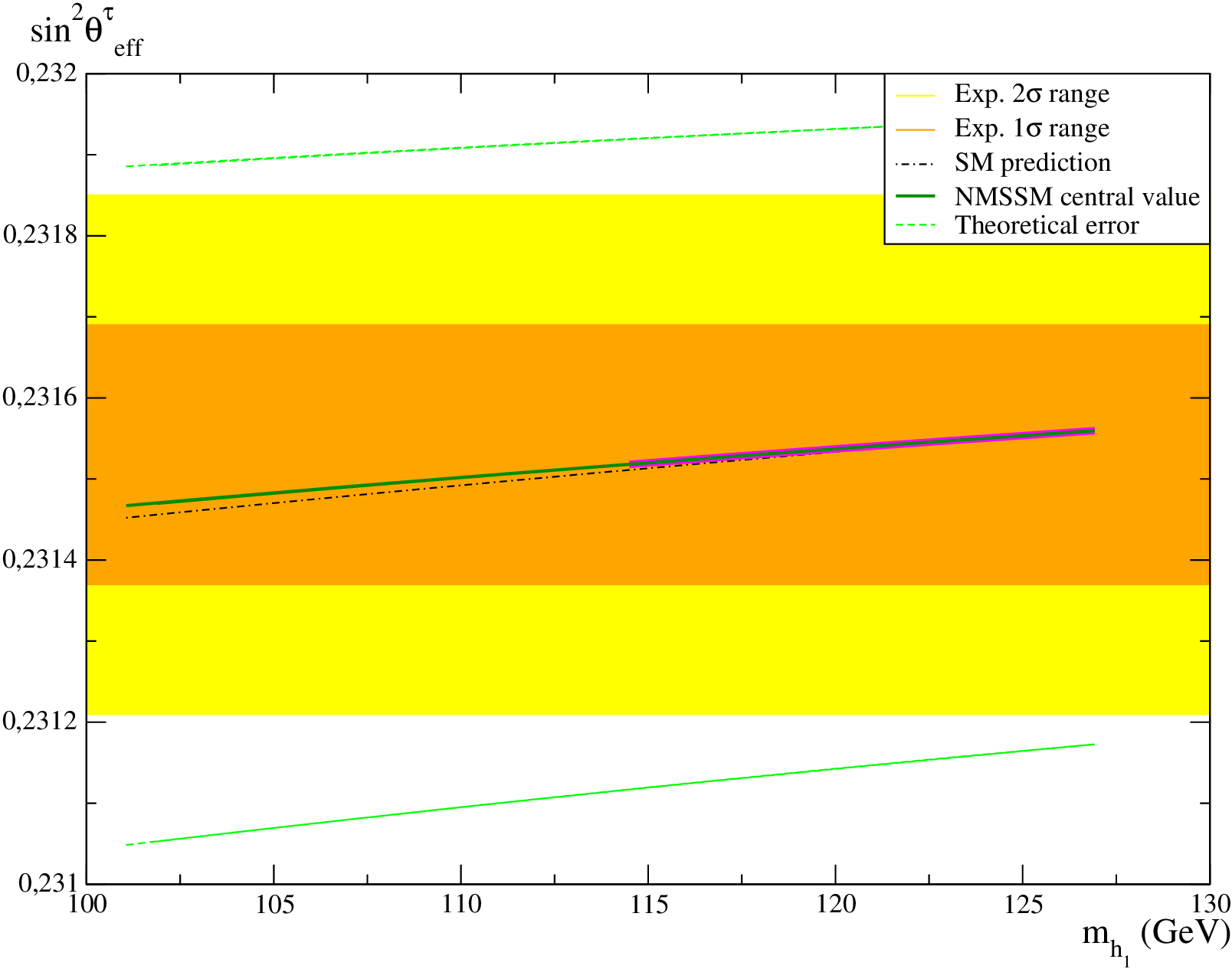}
 \end{center}
\caption{NMSSM prediction for $M_W$ and $\sin^2\theta^{\tau}_{\mbox{\tiny eff}}$ in the low $\tan\beta$ 
regime ($\tan\beta=2$). The supersymmetric sector was chosen heavy ($\sim2$~TeV), in order to highlight Higgs 
effects. The colour-code is similar to {\em e.g.} Fig. \ref{fig:obs(mH+)}.}
\label{fig:obs(mh1lamb)}
\end{figure}

As explained above, we focus on Higgs effects, which, in the MSSM limit, were found dominated by the SM-like 
contribution of the light doublet state. The same remains true for the low $\tan\beta$ NMSSM, as can be 
observed in Fig. \ref{fig:obs(mh1lamb)} (with a heavy SUSY scale $\sim2$~TeV): $M_W$ and $\sin^2\theta^{
\tau}_{\mbox{\tiny eff}}$ are shown in terms of the lightest CP-even Higgs mass (in this case almost purely 
doublet) and found almost identical to the SM-predictions for similar Higgs masses. Note that the low 
$\tan\beta$ effect on Fig. \ref{fig:obs(tanB)}, in the excluded range of the MSSM limit, was simply associated 
with the MSSM light CP-even state becoming light (hence violating LEP constraints): this behaviour 
characterizes in no way that of the low-$\tan\beta$ NMSSM where the light CP-even Higgs mass need also exceed 
$\sim114$~GeV to escape LEP bounds (except in specific scenarii which will be discussed in the following 
subsections). Naturally, lighter SUSY spectra can be associated to this regime, leading to increased $M_W$ as 
we discussed for the MSSM limit.

\subsection{NMSSM Light CP-odd Higgs}
In the NMSSM, light CP-odd Higgs states $A_1$ (as light as a few GeV) represent an interesting phenomenological 
possibility. Such particles have indeed vanishing $V-V-A_1$ ($V=W,Z$) couplings which allows them to circumvent 
most of the direct collider constraints. $B$-physics processes \cite{Bphys}, in particular $\bar{B}_s^0\to\mu^+
\mu^-$ or $\bar{B}_s^0\to X_s\mu^+\mu^-$, and $\Upsilon$ decays \cite{Upsilon}, for $m_{A_1}\leq m_{\Upsilon}$, 
constrain however significantly large values of the reduced-coupling of the $A_1$ to down-type quarks, 
$X_d\equiv\cos\theta_A\,\tan\beta$, where $\cos\theta_A$ quantifies the amount of doublet component in the 
$A_1$. The very-light mass ($m_{A_1}\lsim 5$~GeV) region was shown to be already essentially excluded 
\cite{Andreas:2010ms}. The light NMSSM CP-odd Higgs is often associated with the so-called ``Ideal Higgs 
scenario'', which we will discuss in the next subsection and in which case it must satisfy $m_{A_1}<2\,m_B$ to 
forbid kinematically the {\em a priori} dominant $A_1\to b\bar{b}$ decay channel and circumvent bounds from 
CP-even Higgs searches at LEP. Yet, in general, light $A_1$ are not necessarily associated with specifically 
light CP-even Higgs ($h_1$ could be as heavy as $\sim140$~GeV in the NMSSM), hence offer a wider 
phenomenological possibility than the ``Ideal Higgs scenario''. In two limits of the parameter space (R 
symmetry, with vanishing Higgs trilinear soft couplings; Peccei-Quinn symmetry, with $\frac{\kappa}{\lambda}\ll
1$), the $A_1$ is the Nambu-Goldstone boson of an approximate and spontaneously-broken global symmetry, which 
justifies the naturality of its lightness. Note that in such cases, $X_d$ remains in general small and the 
decoupled $A_1$ proves hard to probe experimentally.

\begin{figure}[t]
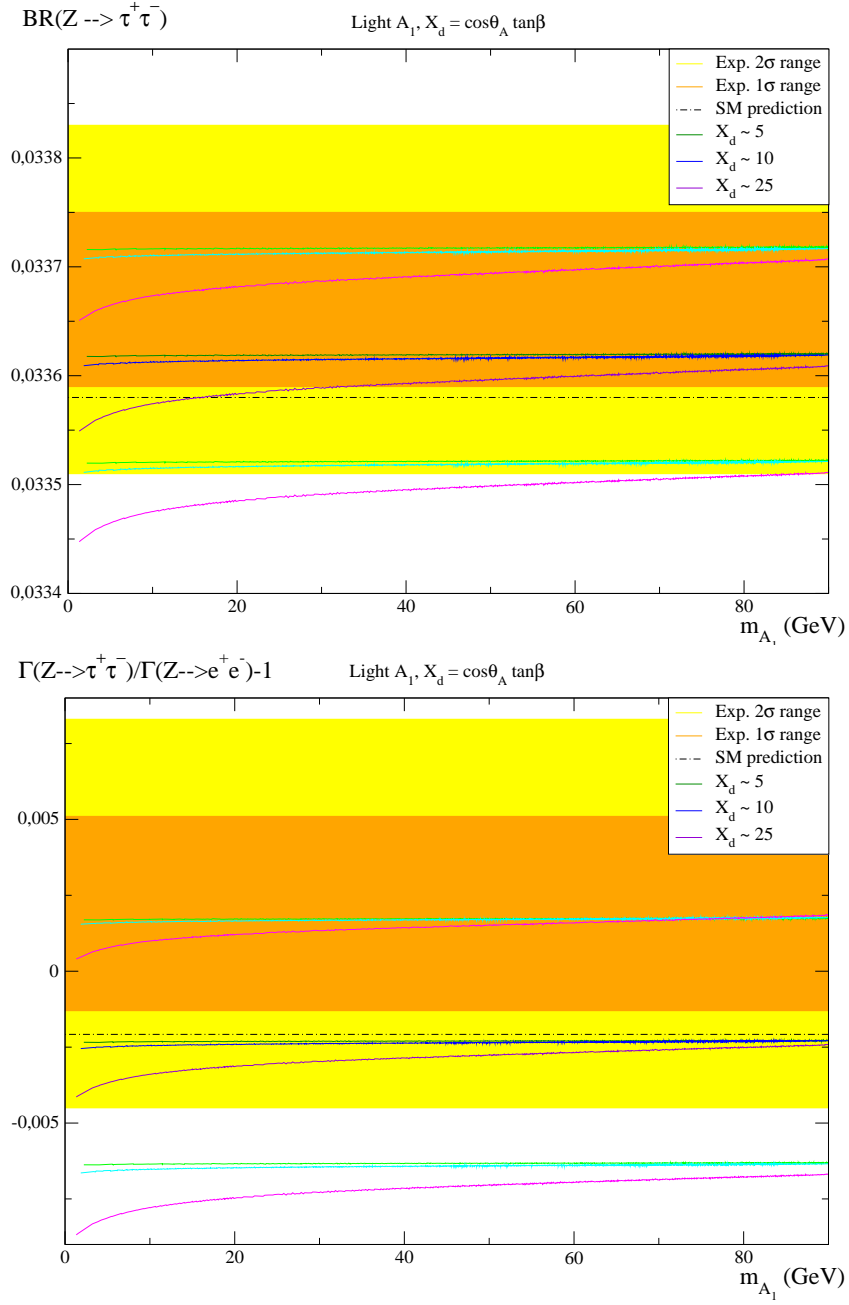

 \begin{center}
\includegraphics[width=11cm]{BRtt_mA1.eps}
\includegraphics[width=11cm]{ratio_mA1.eps}
 \end{center}
\caption{Impact of the NMSSM light CP-odd Higgs $BR(Z\to\tau^+\tau^-)$ and $\Gamma(Z\to\tau^+\tau^-)/
\Gamma(Z\to e^+e^-)-1$. Sfermion effects are suppressed by the choice of a heavy scale $\sim2$~TeV. The 
reduced coupling of the light CP-odd Higgs $X_d\equiv\cos\theta_A\,\tan\beta$ is maintained constant ($\pm0.3$) 
at $5$ (dark green curve), $10$ (dark blue curve) and $25$ (violet curve). The theoretical uncertainty is 
bounded by the paler (green, blue and pink) curves. Corresponding effects on $M_W$ and $\sin^2\theta^{\tau}_{
\mbox{\tiny eff}}$ were found negligible.}
\label{fig:obs(mA1)}
\end{figure}

We investigate the effects of a light CP-odd Higgs on $BR(Z\to\tau^+\tau^-)$ and $\Gamma(Z\to\tau^+\tau^-)/
\Gamma(Z\to e^+e^-)-1$ in Fig. \ref{fig:obs(mA1)} (the effects on $M_W$ and $\sin^2\theta^{\tau}_{
\mbox{\tiny eff}}$ were found negligible). Expectedly, a light CP-odd Higgs can contribute to these observables 
when its reduced-coupling to leptons (and down-type quarks) $X_d$ is enhanced. The corresponding contribution 
is found to worsen the agreement of the theoretical prediction with the experimental limits. Yet, even fairly 
large values of $X_d$ (such as $\sim25$ in Fig. \ref{fig:obs(mA1)}) remain compatible with the current 
precision. In view of the upper-bounds placed on $X_d$ by $B$- and $\Upsilon$-processes \cite{Bphys,Upsilon} 
($X_d\lsim2-3$, for $m_{A_1}\lsim10$~GeV), we may conclude that the light-$A_1$ scenario is consistent with 
leptonic $Z$-decays.

\subsection{Light CP-even doublet-like Higgs}\label{lightdoub}
A widely-studied NMSSM-specific scenario is that of a CP-even (doublet) Higgs decaying unconventionally into a 
pair of light CP-odd Higgs (below the $B-\bar{B}$ threshold), hence escaping LEP constraints from $e^+e^-\to 
Z(h_1\to b\bar{b})$, so that $m_{h_1}\sim90-100$~GeV remains phenomenologically viable. It was even argued that 
$m_{h_1}\sim98$~GeV could be used as an interpretation of the $2.3\,\sigma$ excess found in  $e^+e^-\to Z
(h_1\to b\bar{b})$, leading to an ``Ideal Higgs scenario'' \cite{Dermisek:2005gg}. The recent {\sc Aleph} 
analysis on $e^+e^-\to Z(h_1\to2A_1\to4\tau)$ \cite{ALEPH:2010aw} seems, however, to constrain efficiently such 
a scenario, although possible ways out, such as reduced $BR(A_1\to\tau^+\tau^-)$ decays for $\tan\beta\sim1-2$
\cite{Dermisek:2010mg}, have been suggested. 
\begin{figure}[t]
 \begin{center}
\includegraphics[width=11cm]{MW_mh1doub.eps}
\includegraphics[width=11cm]{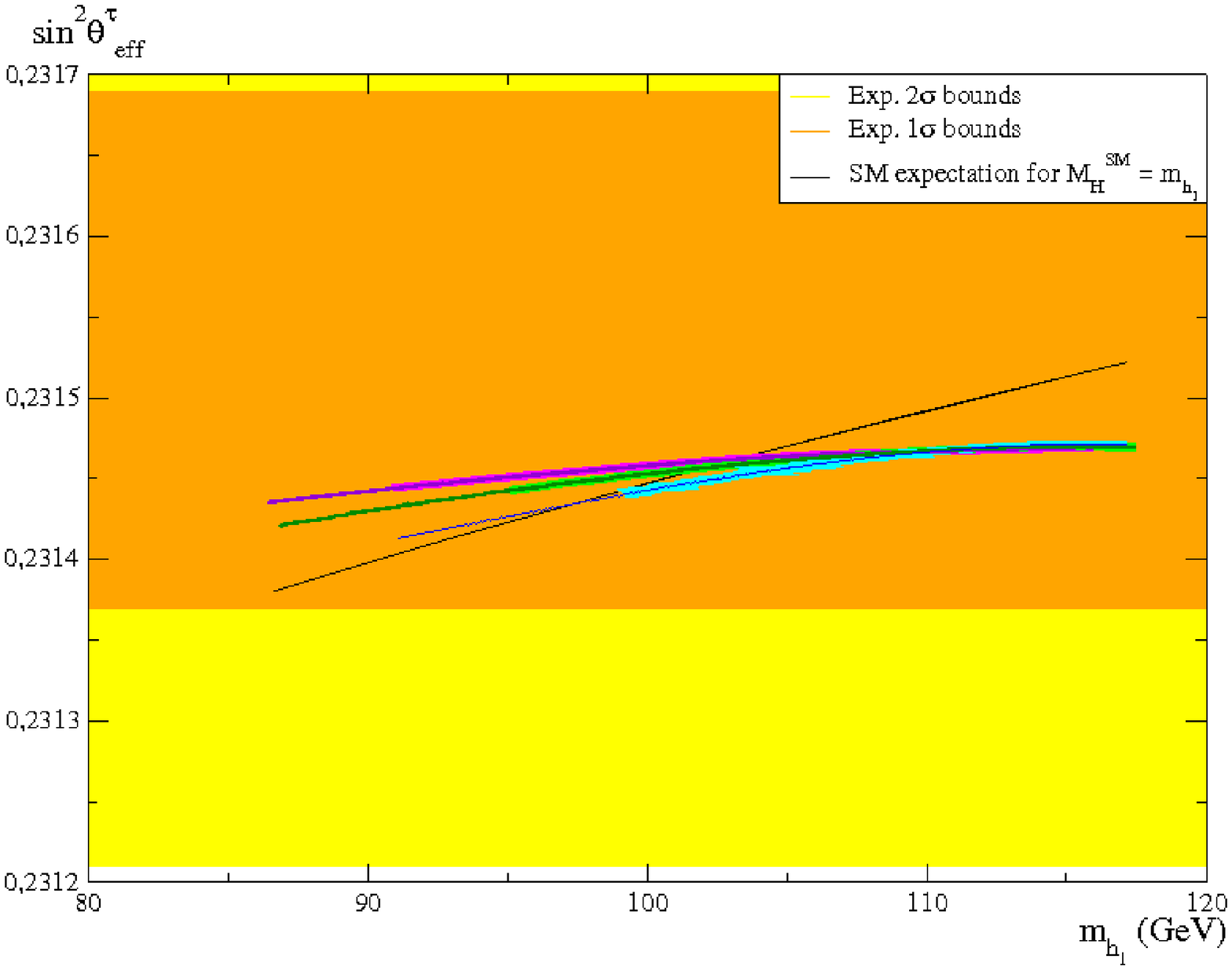}
 \end{center}
\caption{Impact of the light doublet-like CP-even Higgs (in an approximate Peccei-Quinn limit $\frac{\kappa}{
\lambda}=0.1$, with $\lambda=0.5$) on $M_W$ and $\sin^2\theta^{\tau}_{\mbox{\tiny eff}}$. As explained in the 
main part of the text, the light CP-even Higgs $h_1$ possesses a significant subdominant singlet component: 
$\sim45\%$, $\sim30\%$ and $\sim15\%$ respectively for the violet, blue and green curves. The lighter ``aura'' 
indicates as before points which were found in agreement with all constraints implemented in NMSSMTools, 
included the recent {\sc Aleph} constraints. These latter limits are those relevant for the lower $h_1$ mass 
range. Note that the ``SM expectation for $M_H^{SM}=m_{h_1}$'' is shown to allow for the comparison, but 
corresponds to Higgs masses which, in the SM, contrary to the NMSSM, are phenomenologically excluded.}
\label{fig:obs(mh1doub)}
\end{figure}

In the following, we will consider a specific realization of this light doublet-Higgs scenario where 
{\sc Aleph} bounds may also be circumvented. An approximate Peccei-Quinn symmetry (characterized by 
$\kappa/\lambda\sim O(10^{-1})\ll1$) protects the light CP-odd Higgs mass (although the R symmetry is more 
frequently used), hence allowing for the unconventional channel $h_1\to2A_1$ (with $m_{A_1}<2m_B$), while 
$\lambda\sim0.5$ is relatively large (hence allowing for specific NMSSM effects). In this setup, the CP-odd 
Higgs is essentially singlet and its effect on $M_W$ can be neglected. The two CP-even light states are 
essentially the doublet MSSM-like state and the singlet component. Due to the large $\lambda$, they can mix 
efficiently (while the heavy doublet decouples). Therefore, the lightest CP-even state, albeit dominantly 
doublet, carries a non-(necessarily) negligible singlet component. This leads, in turn, to a reduced production 
cross-section. We found realizations of this scenario for $\tan\beta$ in the range $2-12$. Three factors 
``conspire'' to circumvent LEP and in particular the {\sc Aleph} limits in this case:
\begin{itemize}
 \item The unconventional $h_1$ decay $h_1\to2A_1$ is the dominant channel and explains the invisibility of 
$h_1$ in the $b\bar{b}$ channel at LEP. Note however that $BR(h_1\to2A_1)$ need not be exactly $1$ ($\sim0.8$ 
still guarantees the invisibility of $h_1$ in $b\bar{b}$ decays, for $m_{h_1}\sim95-100$~GeV).
 \item The production cross-section $e^+e^-\to Z h_1$ is slightly reduced (with respect to the SM 
cross-section), due to the small singlet component of $h_1$ (by at least $\sim 10\%$).
 \item $BR(A_1\to\tau^+\tau^-)\sim0.75-0.9<1$ (for $m_{A_1}\sim9-10$~GeV), due to secondary decay channels 
(in particular $A_1\to gg$).
\end{itemize}
We choose a significantly heavy sfermion sector $m_{\mbox{\tiny Sf.}}=1.2~\mbox{TeV}=-A_{\mbox{\tiny Sf.}}$ and 
a hierarchical gaugino structure $6M_1=3M_2=M_3=1.2$~TeV. We checked numerically that the contribution of 
supersymmetric particles remained approximately constant in comparison to the Higgs effects. The light CP-odd 
Higgs mass is kept in the range $9-10$~GeV. Being almost entirely singlet, it contributes negligibly to the 
considered observables.

The variations of $M_W$ (and $\sin^2\theta^{\tau}_{\mbox{\tiny eff}}$) as a function of the mass of the light 
CP-even state, in the range $\sim[80,120]$~GeV, are shown in Fig. \ref{fig:obs(mh1doub)}, for fixed amounts of 
the subdominant singlet component $S_{13}^2\sim0.15,\ 0.30,\ 0.45$ ($\pm\,5\cdot10^{-3}$). We observe that the 
effect, even though the light state is dominantly doublet, remains comparatively smaller than that of a  SM 
Higgs in the corresponding mass range (were such masses still allowed for the SM). This observation may be 
easily understood. In our scan, the MSSM-like diagonal light Higgs mass (noted as $m_{h^0}$; note that this 
quantity still benefits from a NMSSM mass contribution $\propto\lambda\mbox{\em v}\,\sin{2\beta}$, as discussed 
in subsection \ref{subsec:largelamb}) is kept approximately constant (since the SUSY masses do not vary). 
Consequently, for a fixed singlet component $S_{13}^2$, the lowering of $m_{h_1}$ is associated to an 
increasing singlet-doublet mixing entry and a heavier second-lighter (singlet-like) state: the more efficient 
contribution of a lighter $h_1$ to $M_W$ is balanced by a decreased $h_2$ contribution (with a $\gsim15\%$ 
doublet component). More quantitatively, whenever the propagators of the two light Higgs states intervene, 
{\em e.g.} in a gauge boson self-energy $\Sigma_V$, the efficient part of the Higgs couplings originates from 
their doublet ``MSSM'' component, that is $g_{\mbox{\tiny NMSSM}}^{h_i}\propto\sqrt{1-S_{i3}^2}\
g_{\mbox{\tiny MSSM}}^{h^0}$. Therefore, with $1-S_{23}^2\sim S_{13}^2$ (since the singlet is essentially 
distributed among the two light states; $q$ denotes the 4-momentum of the considered Higgs-line):
\begin{equation}
 \Sigma_V^{\{h_1,h_2\}}\propto\left[\frac{1-S_{13}^2}{q^2-m_{h_1}^2}+\frac{S_{13}^2}{q^2-m_{h_2}^2}\right]
(g_{\mbox{\tiny MSSM}}^{h^0})^2
\end{equation}
However, the diagonalization of the MSSM-like and singlet states gives $m_{h^0}^2\sim(1-S_{13}^2)\,m_{h_1}^2
+S_{13}^2\,m_{h_2}^2$, so that $(1-S_{13}^2)[q^2-m_{h_1}^2]^{-1}+S_{13}^2[q^2-m_{h_2}^2]^{-1}\sim[q^2-
m_{h^0}^2]^{-1}$, and we obtain that the Higgs contribution is approximately given by that of its MSSM-like 
diagonal component. The latter, however, is kept constant in our case, and cannot be significantly lowered 
anyway, since {\sc Aleph} constraints would exclude a pure doublet state for $m_{h_1}\lsim103-105$~GeV 
(depending on the details of  $BR(A_1\to\tau^+\tau^-)$ and  $BR(h_1\to A_1A_1)$; note also that small 
$\tan\beta\sim1-2$ lead to reduced $BR(A_1\to\tau^+\tau^-)$, allowing for even lighter doublet-states 
\cite{Dermisek:2010mg}).
Therefore, the NMSSM-Higgs effect we obtain is reduced and, in Fig. \ref{fig:obs(mh1doub)}, the generated shift 
in $M_W$ is only of a few MeV in the authorized $m_{h_1}$ range (not violating {\sc Aleph} constraints).
\begin{figure}[t]
 \begin{center}
\includegraphics[width=11cm]{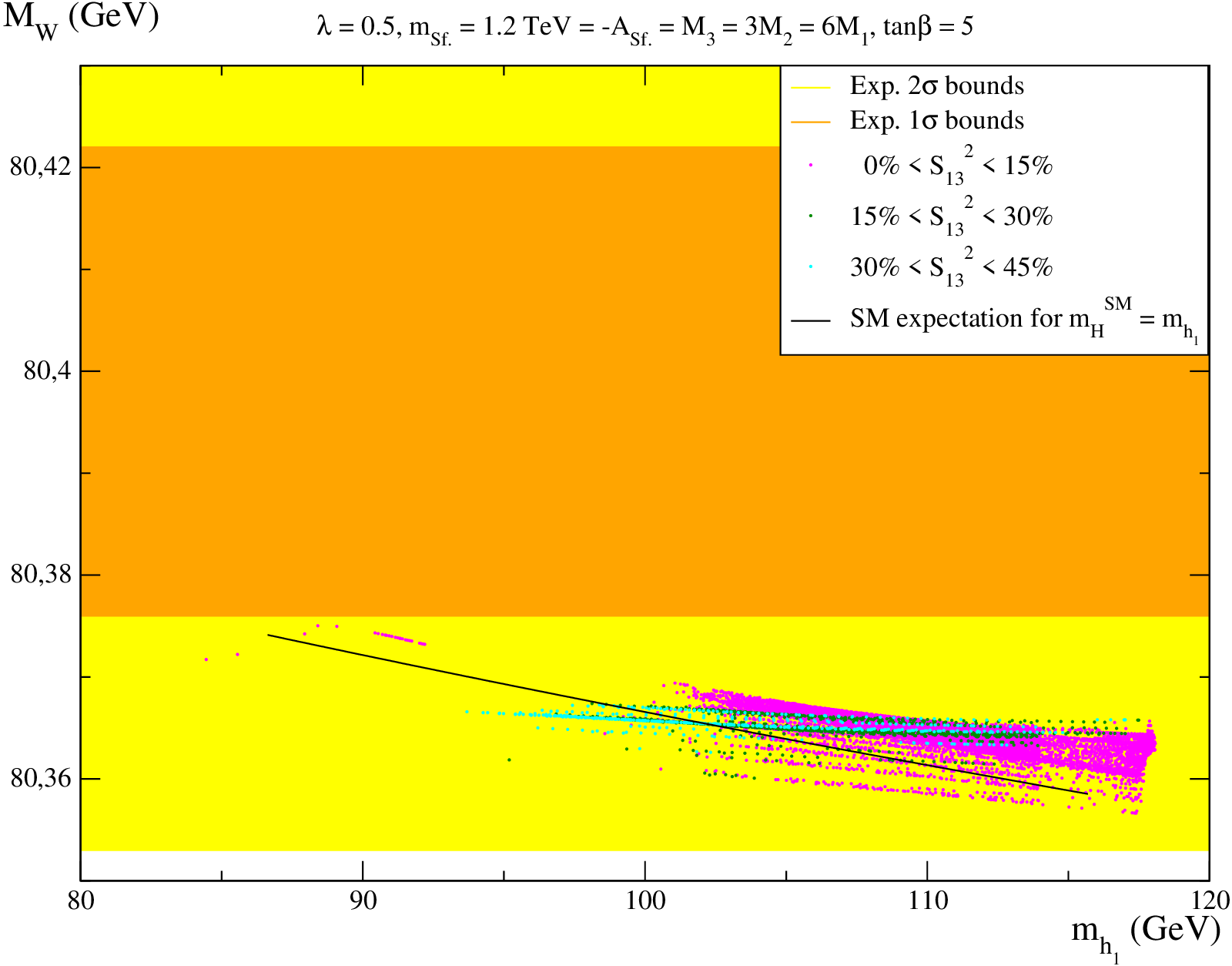}
 \end{center}
\caption{Impact of the light doublet-like CP-even Higgs (with $\lambda=0.5$, $\tan\beta=5$) on $M_W$. This 
scatter plot is obtained by scanning over the parameters $A_{\lambda}$, $A_{\kappa}$, $\mu_{\mbox{\tiny eff}}$ 
and $\kappa$ (with the requirement that $m_{A_1}=9-10$~GeV) and retaining all the points allowed by the 
phenomenological constraints implemented in NMSSMTools.}
\label{fig:MW(scandoub)}
\end{figure}
\begin{figure}[t]
 \begin{center}
\includegraphics[width=11cm]{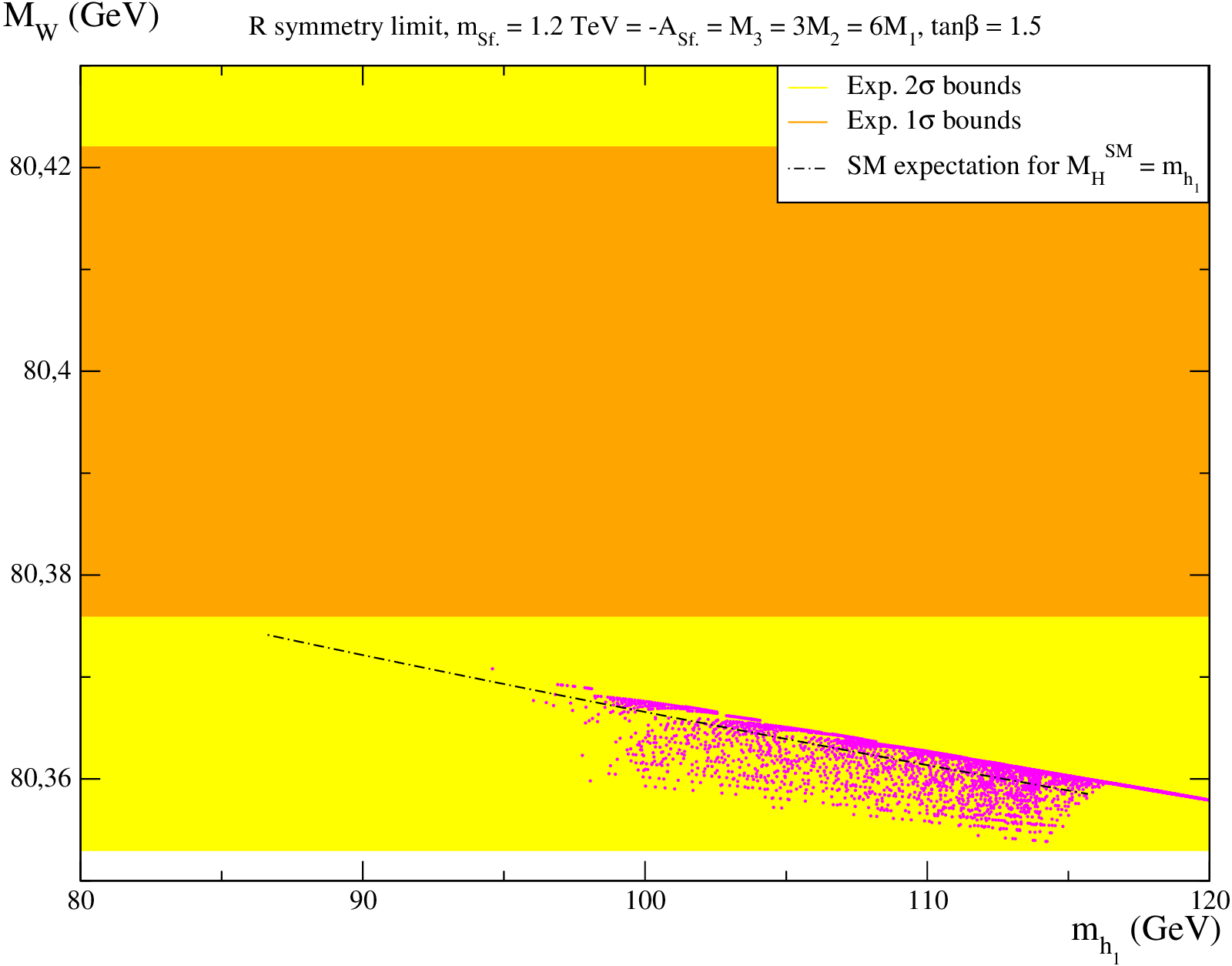}
 \end{center}
\caption{Impact of the light doublet-like CP-even Higgs (with $\lambda=0.5$, $\tan\beta=1.5$) on $M_W$. This 
scatter plot is obtained by scanning over the parameters $A_{\lambda}$, $A_{\kappa}$ ($|A_{\lambda,\kappa}|
<50$~GeV to ensure an approximate R-symmetry), $\mu_{\mbox{\tiny eff}}$ and $\kappa$ (with the requirement that 
$m_{A_1}=9-10$~GeV) and retaining all the points allowed by the phenomenological constraints implemented in 
NMSSMTools.}
\label{fig:MW(scanRdoub)}
\end{figure}

Note however that we chose here a very specific realization of the unconventionally-decaying Higgs scenario, 
and that slightly larger effects might be generated by relaxing {\em e.g.} the Peccei-Quinn symmetry 
requirement. This is illustrated by Fig. \ref{fig:MW(scandoub)}, where we study a scatter plot obtained by 
varying all the parameters of the Higgs sector ($A_{\lambda}$, $A_{\kappa}$, $\mu_{\mbox{\tiny eff}}$ and 
$\kappa$) with the exception of $\lambda=0.5$ and $\tan\beta=5$. Note that as we depart from the Peccei-Quinn 
limit, points with large $S_{13}^2>15\%$ become scarce (because the diagonal singlet mass, no longer protected 
by small $\kappa/\lambda$, tends to be large and the CP-even singlet cannot couple as efficiently with the 
doublet). The pink points correspond to almost entirely doublet $h_1$ states (so that $m_{h_1}\sim m_{h^0}$) 
and we observe, expectedly, that they are scattered around the SM prediction for $M_H^{SM}=m_{h_1}$ (which is, 
in the SM, excluded by LEP). Although slightly less efficient, points with larger $S_{13}^2$ give comparable 
results.

Finally, we extend our analysis to the region $\tan\beta=1.5$, with an approximate R-symmetry (which we 
implement by the requirement $|A_{\lambda,\kappa}|<50$~GeV) protecting the light CP-odd mass. $\lambda=0.5$ 
continues to ensure the possibility of specific NMSSM effects. The {\sc Aleph} limits are circumvented by the 
reduced $BR(A_1\to\tau^+\tau^-)$ (at low $\tan\beta$ \cite{Dermisek:2010mg}), although a small $S_{13}^2$ 
($\lsim8\%$) also often intervenes. In the scatter plot of Fig. \ref{fig:MW(scanRdoub)}, we scan again over 
$A_{\lambda}$, $A_{\kappa}$, $\mu_{\mbox{\tiny eff}}$ and $\kappa$ (note that the allowed points we obtain all 
verify $\kappa\sim0.3-0.5$). The allowed points again cluster around the (excluded by LEP) SM central value for 
$M_H^{SM}=m_{h_1}$.

As a conclusion for this subsection, we find that the NMSSM scenario with light CP-even doublet-like Higgs in 
the mass-range $[95-115]$~GeV allows for a specific contribution, similar to that a doublet SM or MSSM 
Higgs would have generated, were such low masses still phenomenologically allowed in these models. Note however 
that what mass seems determinant is the pure doublet (MSSM-like) mass $m_{h^0}$ and not, in case of a mixing 
with the singlet, that of the light eigenstate $m_{h_1}$. Note again that in this analysis, we tried to 
isolate this specific Higgs effect by suppressing the contribution of supersymmetric particles 
($m_{\mbox{\tiny Sf.}}=1.2$~TeV). In the general case however, SUSY effects also intervene and add to the 
Higgs effect.

\subsection{Light CP-even singlet-like Higgs}
\begin{figure}[t]
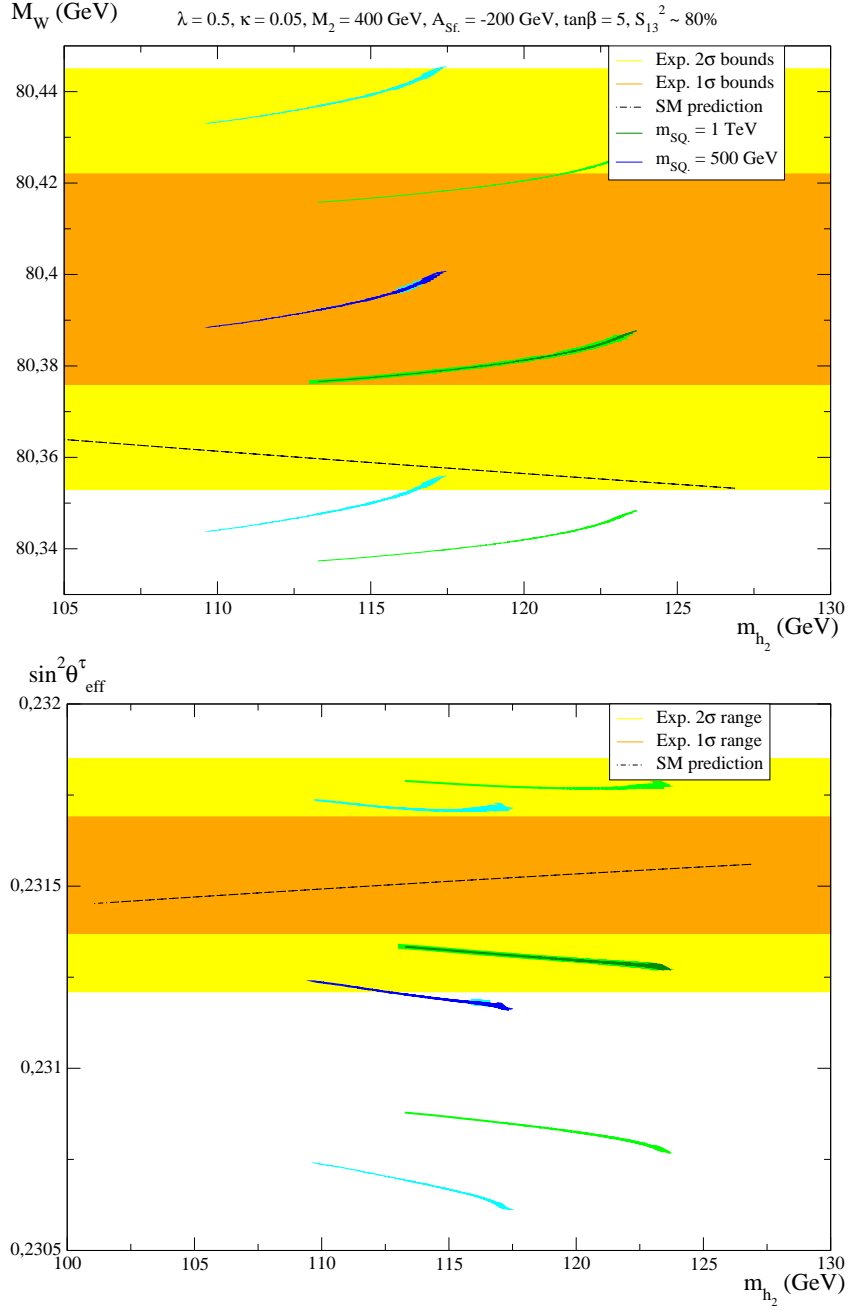

 \begin{center}
\includegraphics[width=11cm]{MW_mh1sing.eps}
\includegraphics[width=11cm]{s2tweff_mh1sing.eps}
 \end{center}
\caption{Light singlet-like CP-even Higgs scenario (in an approximate Peccei-Quinn limit $\frac{\kappa}{
\lambda}=0.1$, with $\lambda=0.5$): $M_W$ and $\sin^2\theta^{\tau}_{\mbox{\tiny eff}}$ are plotted against the 
mass of the observable doublet-like state $m_{h_2}$, for $S_{13}^2\sim0.8\,(\pm0.002)$ (note that the 
corresponding $h_1$ masses, for the allowed points, are in the range $\sim[40,110]$~GeV). The dark-green and 
dark-blue curves correspond to the theoretical central values for universal squark soft-masses of respectively 
$1$~TeV and $500$~GeV (The lighter ``aura'' indicates points which were found in agreement with all constraints 
implemented in NMSSMTools). The error bars are displayed in light-green and light-blue respectively.}
\label{fig:obs(mh1sing)}
\end{figure}

In the following, we consider another popular Higgs scenario, where the lightest CP-even Higgs is singlet-like, 
hence circumvents LEP constraints in virtue of a reduced production cross-section in $e^+e^-$ collision. For a 
purely decoupled singlet, this particle can be in principle as light as a few $\sim10$~GeV but also has 
negligible effects on the doublet sector. We will thus rather focus on CP-even singlets with a sizable doublet 
component: we shall use $S_{13}^2\sim80\%$ in Fig. \ref{fig:obs(mh1sing)}, which satisfies LEP constraints provided 
$m_{h_1}\gsim40$~GeV. In this case, the singlet-doublet mixing uplifts the mass of the light doublet-like Higgs 
state $h_2$, with respect to the pure doublet mass $m_{h^0}$ (due to the splitting effect).

We obtain a specific realization of this scenario in an approximate Peccei-Quinn limit of the parameter space 
$\kappa/\lambda=10^{-1}\ll1$, $\lambda\sim0.5$: the setup is comparable to that of subsection \ref{lightdoub}, 
but with an inverted hierarchy between the singlet and light-doublet states. Note that, neglecting the 
Peccei-Quinn violating terms, the tree-level diagonal squared mass of the singlet component is given by 
$m_{S^0}^2\simeq\frac{\lambda^2\mbox{\small\em v}^2 A_{\lambda}\sin{2\beta}}{2\mu_{\mbox{\tiny eff}}}$ and is 
naturally at the electroweak scale or smaller (since $\sin{2\beta}\ll1$ for $\tan\beta\gg1$). Moreover, the 
singlet and light-doublet components can couple efficiently in the large $\lambda$ regime so that significant
mixing may be easily achieved.
\begin{figure}[t]
 \begin{center}
\includegraphics[width=11cm]{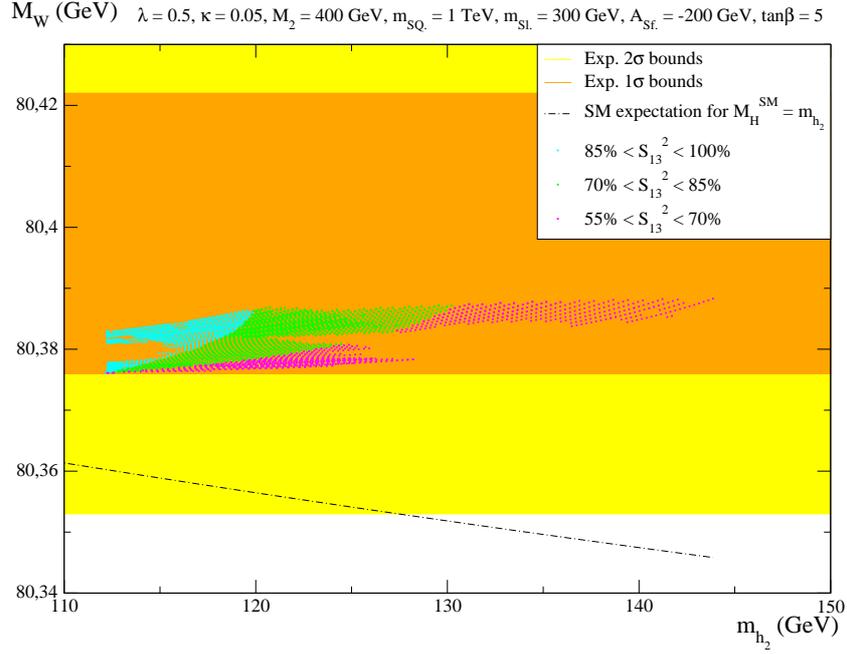}
 \end{center}
\caption{Light singlet-like CP-even Higgs scenario (in an approximate Peccei-Quinn limit $\frac{\kappa}{
\lambda}=0.1$, with $\lambda=0.5$): scatter plot for $M_W$ against the mass of the observable doublet-like 
state $m_{h_2}$. Squarks are taken at a heavy scale $m_{\tilde{Q}}=1$~TeV, while, for sleptons, $m_{
\mbox{\tiny Sl.}}=300$~GeV. Light blue, green and pink points satisfy the constraints implemented in NMSSMTools 
and verify respectively $S_{13}^2\in[0.85,1]$, $\in[0.7,0.85]$, $\in[0.55,0.7]$. Corresponding $h_1$ masses are
in the range $[40,110]$~GeV. The dot-dash black curve shows the expected SM central value for a SM 
Higgs of similar mass. The ``cuts'' inside the allowed regions are essentially related to LEP limits on Higgs 
production.}
\label{fig:MW(scansing)}
\end{figure}

In this setup, several remarkable effects may influence EWPOs: $h_1$ is significantly lighter than LEP-bounds 
for doublet states and carries a sizable doublet component ($\sim20\%$ in Fig. \ref{fig:obs(mh1sing)}); the 
mass of the ``visible'' light-doublet state receives a contribution from the singlet-doublet mixing, hence 
allows for reduced pure doublet masses $m_{h^0}$; for the same reason, squark masses may be smaller, hence 
contribute more to $M_W$. This discussion is illustrated by Fig. \ref{fig:obs(mh1sing)}, showing $M_W$ and 
$\sin^2\theta_{\mbox{\tiny eff}}^{\tau}$ in terms of the mass of the observable Higgs state, $m_{h_2}$. Note 
that $M_W$ now increases with $m_{h_2}$: as the $h_2$ becomes heavy, the $h_1$ state, carrying a $20\%$ doublet 
component, becomes correspondingly light, due to the splitting effect between the light-doublet and singlet 
components, hence has an enhanced contribution to $M_W$. However, excessively light $h_1$ ($m_{h_1}\lsim40$~GeV 
for $S_{13}^2\sim0.8$) are eventually constrained by LEP searches. Lighter sfermion spectra yield both an 
increased SUSY contribution and a reduced diagonal doublet mass. This last property implies that (for equivalent 
$h_1$ masses) $h_2$ is also lighter. The limiting factor for light sfermions in Fig. \ref{fig:obs(mh1sing)}, 
however, was found to be the stability of the electroweak-symmetry breaking vacuum. As the agreement of $M_W$ 
with its experimental measurement is improved, that of $\sin^2\theta_{\mbox{\tiny eff}}^{\tau}$ again worsens 
correspondingly, albeit still within acceptable proportions with respect to the error bars.

In Fig. \ref{fig:MW(scansing)}, we show a scatter plot (over $A_{\lambda}$, $A_{\kappa}$ and 
$\mu_{\mbox{\tiny eff}}$) for $\lambda=0.5$, $\kappa/\lambda=0.1$ and a dominantly singlet light CP-even Higgs 
state. We observe that Higgs effects may generate a large $M_W$, even though the observable Higgs is as heavy 
as $\sim140$~GeV. We stress that this effect is allowed in the NMSSM due essentially to the interplay between 
doublet and singlet CP-even states and is thence NMSSM-specific.

\section{Conclusions}
We have studied $M_W$ and other electroweak observables associated with $Z$ leptonic decays in the 
NMSSM. Similarly to the MSSM, most of the acceptable parameter space of this model is consistent with the 
present accuracy available for the corresponding observables. Such an agreement with respect to EWPOs shows 
the robustness of supersymmetry-inspired extensions of the SM with respect to the description of the 
electroweak-symmetry breaking.

As a general trend, new-physics (N/MSSM) effects are found to improve the theoretical prediction for $M_W$ with 
respect to the SM (which however remains compatible with existing bounds), the leading new contribution 
originating from (moderately) light supersymmetric spectra ($\sim300-500$~GeV). Specific effects in the NMSSM 
are associated with its extended Higgs sector. Yet, most of the Higgs effects in the electroweak precision 
observables under consideration are related to the mass of the light CP-even doublet component, lying in the 
$\sim[90,140]$~GeV range in the NMSSM. Corresponding effects remain generally small with respect to the current 
precision, although fluctuations of $M_W$ by a few MeV can be generated in scenarios with very light doublet-like 
states or efficient doublet-singlet mixings, as soon as a state with non-negligible (a few $10\%$) doublet 
component becomes light. In the presence of a heavy supersymmetric spectrum, the unconventional Higgs spectrum 
of the NMSSM could thus relieve the tension in $M_W$, even for an apparently-heavy observable Higgs boson.

The possible presence of a light CP-odd Higgs in the NMSSM spectrum may also affect the universality of 
leptonic $Z$-decays. Note that corresponding effects tend to worsen (albeit below significance) the agreement 
with the experimental values.  Nevertheless, only large (unrealistic) $A_1\bar{l}l$ couplings would lead to a 
significant deviation, which, in view of existing bounds (and naturalness considerations), are unlikely. The 
possibility of light CP-odd states and, in particular, their involvement in specific NMSSM scenarii (through 
unconventional decays) thus remain essentially consistent with the observables under consideration.

Further information might be extracted from hadronic $Z$-decays, although existing studies \cite{Cao:2008rc} 
already showed that no improvement was to be expected in the $Z\to b\bar{b}$ asymmetry. Improved experimental 
precision may strengthen the relevance of these EWPOs in order to constrain supersymmetry-inspired models in 
the future (and possibly discriminate among them), reducing both the experimental uncertainty and the 
theoretical parametric error. Yet much effort will then be necessary to reduce the higher-order uncertainty and 
thus achieve theoretical results of comparable accuracy. 

\section*{Acknowledgements}
The authors acknowledge discussions with S. Heinemeyer and G. Weiglein and also wish to thank U. Ellwanger and 
U. Nierste for useful discussions and comments. The work of F.D. was supported by the EU Contract No. 
MRTN-CT-2006-035482, FLAVIAnet, by DFG through project C6 of the collective research centre SFB-TR9 and by 
BMBF grant 05H09VKF.

\end{document}